\documentclass[12pt]{article}
\usepackage{graphicx}
\usepackage[section]{placeins}

\begin{document}
\title{A simple atomistic model for the simulation of the gel phase of 
lipid bilayers}

\author{G.~La Penna$^{^{a)}}$ S.~Letardi$^{^{b)}}$ V.~Minicozzi$^{^{c)}}$\\
S.~Morante$^{^{c)d)*)}}$ G.C.~Rossi$^{^{c)e)}}$ G.~Salina$^{^{e)}}$\\\\
\small $^{a)}$Istituto di Studi Chimico-Fisici di Macromolecole Sintetiche e
Naturali, CNR \\ \small Via De Marini 6, 16149 Genova, Italy \\
\small $^{b)}$ENEA, Casaccia \\
\small $^{c)}$Dipartimento di Fisica, Universit\`a di Roma {\it Tor Vergata}
\\
%\small Via della Ricerca Scientifica, 00133 Roma, Italy \\
\small $^{d)}$INFM, Unit\`a di Roma 2 \\
\small $^{e)}$INFN, Sezione di Roma 2 \\
%\small Via della Ricerca Scientifica, 00133 Roma, Italy \\
\small Via della Ricerca Scientifica, 00133 Roma, Italy \\
\medskip}

\maketitle
\vspace{3truecm}
\small $^{*)}${\it Correspondence to:} Silvia Morante, Dipartimento di Fisica,
Universit\`a degli Studi di Roma ``{\it Tor Vergata}", Via della Ricerca
Scientifica, 00133 Roma, Italy (Phone Number: +39-0672594554;
FAX Number: +39-062025259; E-mail address: morante@roma2.infn.it)

\vfill

\newpage
\begin{abstract}
{In this paper we present the results of a large-scale numerical
investigation of structural properties of a model of cell membrane,
simulated as a bilayer of flexible molecules in vacuum. The study was
performed by carrying out extensive Molecular Dynamics simulations, in the
($NVE$) {\it micro-canonical ensemble}, of two systems of different sizes
($2\times 32$ and $2\times 256$ molecules), over a fairly large set of
temperatures and densities, using parallel platforms and more standard
serial computers. Depending on the dimension of the system, the dynamics
was followed for physical times that go from few hundred of picoseconds for
the largest system to 5--10 nanoseconds for the smallest one. We find that
the bilayer remains stable even in the absence of water and neglecting
Coulomb interactions in the whole range of temperatures and densities we
have investigated. The extension of the region of physical parameters that
we have explored has allowed us to study significant points in the phase
diagram of the bilayer and to expose marked structural changes as 
density and temperature are varied, which are interpreted as 
the system passing from a crystal to a gel phase.}
\end{abstract}
\section{Introduction}
\label{intro}
The simulation of realistic models of cell membranes is of the utmost
importance, if not for immediate medical use, certainly for the
development of new conceptual and practical tools in the application of
Molecular Dynamics (MD) methods to these and similarly complex systems. A
lot of research activity has gone in this direction (see for
instance~\cite{HSS,K,TTK,KLEIN_DPPC,BEB,ZXMT} and references therein), but
we are still far from having a complete understanding of the dynamic and
thermodynamic properties of this kind of systems, not to mention the
problem of simulating the formation and the dynamics of pores and ion
channels (see, however, the recent very interesting work of ref.~\cite{DIECK}).

To predict the structure of molecular layers in different
physico-chemical conditions by computer simulations, 
several modeling strategies have been proposed and investigated.
Among them we would like to mention the following.

1) Atomistic detailed approaches in which bilayer constituent molecules,
water and possible counter-ions are modeled explicitly in full detail. The
interaction between electrostatic point charges are 
evaluated through Ewald related summation techniques, while the bias due to
the use of periodic boundary conditions is reduced by performing
simulations in extended {\it ensembles}. Recent examples of this kind of
investigations can be found in refs.~\cite{TTK,KLEIN_DPPC,PASTOR}.

2) Simplified models of structures formed by surfactants have
been constructed that retain some of the essential interactions between
the constituent particles, like the Lennard-Jones (LJ) forces
among the hydrophobic tails and between solvent and hydrophilic
molecular heads. On top of them an {\it{ad hoc}} repulsive soft core 
describing the interaction between hydrophobic and hydrophilic 
particles is added to allow self-assembling of micelles 
and bilayers~\cite{GOETZ}.

3) Models in which only a single molecule or chain is modeled
explicitly in terms of a realistic all-atom description, while 
the effect of the remainder of the system is parametrized by a mean-field
deterministic force plus appropriate random forces. Models of this type
are successfully treated by Monte Carlo simulations and/or stochastic
techniques. In this context we wish to recall the fairly good 
agreement with experiments obtained in the evaluation of the NMR
deuterium order parameters in the case of lipid 
bilayers, when Brownian Dynamics is used for their 
simulations~\cite{KARPLUS}.

4) In the field of liquid crystals, where certain properties of layers
often can be explored only through computer simulations, even more
simplified force-fields, like that proposed by Gay and Berne~\cite{GAY} (GB
hereafter) with the successive modifications of refs.~\cite{ZEWDIE}
and~\cite{LUCK}, are used. Despite their simplicity these models allow the
exploration and the study of various phases of the system, including smectic
phases characterized by a number of different tilt angles. 

Comparing different models and trying to assess the effects of the
various contributions that are included or neglected in different
instances, is not a simple task. An example of this situation is the
comparison between the GB potential~\cite{GAY} and its closest LJ
representation. The GB potential was originally devised so as
to mimic the interaction between two identical linear arrays
composed by four LJ sites each. The properties of a fluid made of
GB particles turn out to be rather different from its LJ
counterpart: the GB fluid has a phase diagram with a well
characterized nematic-isotropic transition temperature, while
apparently the LJ counterpart does not show any phase transition of this
kind when only the temperature is varied. This is a case where the
use of a simplified potential, far from leading to an
impoverishment of the model, turns out to give rise to an unexpected and
very interesting thermodynamic behaviour.

In the case of lipid molecules the comparison among different 
modeling strategies are much more difficult because of the
occurrence of different types of interaction sites (head,
tail and solvent) and because of the special importance of the
internal molecular degrees of freedom, that may greatly influence the
inter-molecular interactions, as well as the related hydrophobic tail
packing and bilayer stability properties.

\begin{figure}[!htbp] 
\centering
\includegraphics[width=8cm]{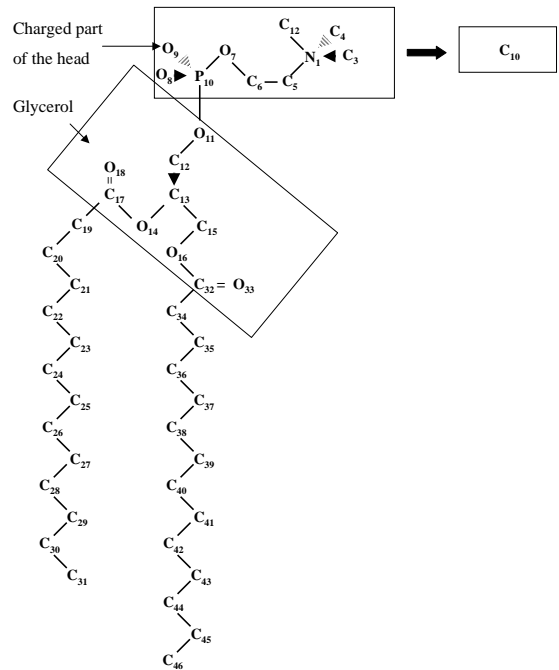}
\caption{The chemical structure of the DMPC molecule and the 
substitution which transforms it into what we have called DMMG. 
Hydrogens are not drawn. Carbons in the CH$_n$ groups are labeled 
progressively in the standard way.}
\label{fig:schema1}
\end{figure}

The purpose of this paper is to perform MD simulations of a
simplified, but still sufficiently rich model of molecular bilayer 
in vacuum. The bilayer constituents are branched molecules composed
by LJ sites with the topology and the internal flexibility of the two
fatty acid tails and glycerol group of the {\it
Dimyristoyl-phosphatidylcholine} (DMPC hereafter) molecule. In
the absence of water, the charged part of the DMPC head has been
replaced by a methyl group (see Fig.~\ref{fig:schema1}).

\vskip 2cm
%***************************************************************************
%***************************************************************************

With respect to an all-atom description of DMPC bilayers~\cite{ZXMT},
we retain in the model only the LJ interactions between the most
hydrophobic portions of the lipid molecules together their
intra-molecular flexibility. These simplifications allow us to single
out and investigate the role of repulsive-dispersive interactions in
the structure of the bilayer, without the complications arising when
electrostatic interactions among heads and solvent are present,
which substantially limit the length of the simulated trajectories. 

Owing to these circumstances we are able to generate very long
trajectories (up to $\sim$10~ns) of sufficiently large systems, which
allow us not only to expose and control certain effects related to
the slow equilibration of the system (which would be important sources of
systematic errors in the analysis of simulations not sufficiently longer
than 1~ns), but also to accurately check the physical consistency
of the parameters of the employed intra-molecular potential.

We notice that in the simplified approach we present in this paper
the bilayer components are not really amphiphilic. This is not, however, a
completely silly approximation. We have checked, in fact, that the
system configurations, occurring at the lowest density we have explored,
in which it happens that some of the molecule tails are turned
up-side-down, have approximately the same potential
energy as the physical situation in which the tails of the
molecules of the two layers are in contact. We find instead
that at the surface density of both the crystal and gel phases the
energy barrier which should be overcome by a molecule to rotate up-side-down
by $180^\circ$~degrees is sufficiently high to prevent this event to
occur in the whole range of temperatures we have explored. This
observation makes us confident that the predictions we have for the
structural properties of the system in these two phases are fully
reliable. As we said the situation, 
is completely different at our lowest density,
which corresponds to that of the liquid-crystal phase.
Evidently in this case the potential barrier is not sufficiently high, 
as we witness the
crossing of it by some of the molecules. Despite this  unphysical feature, we
remark that our bilayer model remains stable at all the values of
temperature and density we have studied (even at the lowest one), in the
sense that, thanks to the strong attractive interactions between the two
layers, none of the molecules leaves the layers.

We also notice that at low surface density the potential energy of
the whole system is presumably smaller than it would be if the
repulsion between solvent and tails were introduced. In models in
which solvatation effects are included~\cite{PASTOR,GOETZ} low
density configurations are essentially inaccessible, precisely
because of a substantial solvent-tail repulsion. 

Given the crudeness of our bilayer model, we do not expect to be able to
accurately
predict thermodynamic parameters, like the dependence of the surface density
or the magnitude of the stress tensor~\cite{KLEIN_DPPC,PASTOR} as functions of
the temperature, mainly because of the lack of the interactions that govern
the structure of the solvent interface and the absence of terms in the
force-field that would prevent the tendency to isotropization 
at much too low surface densities we have mentioned above.
Therefore, in order to study the physics of the
different phases of our bilayer model, we decided to fix the surface density
at the experimentally estimated values ($NVE$-simulations), rather than
trying to work in an extended {\it ensemble} at fixed (lateral) pressure.

As for the role of water, we tentatively assume that, once the
bilayer is formed in the polar fluid, its structural and
thermodynamic behaviour will be mainly governed by the packing and
flexibility properties of the hydrocarbon tails~\cite{LIBMEM}. Most
of the effects of water on the detailed physico-chemical behaviour
of the membrane are indeed of secondary importance for the kind
of questions we are addressing here. In the picture we have in mind,
in fact, water is only important in that it behaves as a sticky
medium for the lipid heads, which modulates the surface density by
more or less penetrating into the bilayer~\cite{WIBS}. Thus water
penetration is held responsible for the difference one sees in the
increase of disorder in the head region, as compared to what happens
to the tails, only to the extent it affects the surface density of
the bilayer. The different behaviour between heads and tails may in
turn be at the origin of the great sensitivity of the structure of
the phase diagram to molecular composition and tail length. 

Even in the simplified model we are considering here, the known
crystal structure of the glycerol group and fatty acid tails of
DMPC~\cite{DENS} is well accommodated in a local minimum of the 
force-field we are using, at the appropriate surface density. The
comparison of our results with those of more refined models can be
of great help in understanding the relevance of more realistic
modelizations of the molecular components of the system and/or the role of
water.

The MD simulations we have performed extend over a rather fine
grid in the temperature-density plane. Thanks to this resolution
and to the large statistics we have collected, we are able
to draw interesting conclusions on the role of LJ interactions
and internal flexibility on the stability and the structural
properties of the crystal and gel phases of our bilayer model.

The paper is organized as follows. In section 2 we describe the model of
bilayer we have constructed and investigated. In section 3 we
discuss the details of our simulation strategy and in section 4 we
report the most significant results we have obtained with a special
emphasis on the information we have collected concerning structure and
properties of the crystal and gel phases of the system. Conclusions can be
found in section 5.

\section{Simulation set-up}

Schematically, a cell membrane can be described as an almost spherical
bilayer consisting of phospho-lipid molecules which separates the
interior of the cell from the external world. Phospho-lipids are
molecules composed by a hydrophilic head and one or two hydrophobic
tails. These peculiar hydropathicity properties lead to the well known
two-sheet structure of the membrane, in which the hydrophilic heads are
in contact with water, present both outside and inside the cell, while
the hydrophobic tails are more or less tail-to-tail pair-wise aligned.

It appears experimentally that an important parameter governing the 
reaction rate of many  biological processes, taking place inside or in
the near vicinity of a membrane, is its ``permeability"~\cite{LIBMEM}.
From this point of view a membrane can be thought as a system made
out of two layers of a smectic fluid, with a permeability which depends
``critically" on temperature, density, the detailed chemical 
composition of the constituent phospho-lipids, the concentration of
chemicals possibly dispersed in the membrane itself or in the solvent,
etc. 

It is clear that a detailed simulation of the dynamics of the membrane 
of a living cell is just impossible and one has to resort to a number
of simplifications. A fortunate circumstance in this respect is that
it appears experimentally that the nature and the location of the
phase transitions, which control a number of important
physico-chemical properties of the membrane, are essentially related
to the  bulk ordering properties of the hydrophobic
tails~\cite{LIBMEM}. Thus a first step in the direction of simulating a
realistic system, and the one which has been also largely followed in
the literature~\cite{HSS,K,KLEIN_DPPC,ZXMT}, is to take a two sheet
system, each consisting of the largest possible number of lipid  molecules
(compatibly with the available computer power), in presence or even in
absence of water, and proceed to study how the relevant order parameters
behave as functions of temperature and surface density and what are the
structural properties of the system in the different regions of its phase
diagram.

Lacking any realistic theoretical modeling of mesoscopic systems, as
membrane or other molecular aggregates of biological relevance, one must
resort to numerical methods, like MD or Monte Carlo and stochastic
simulations of various kinds, in order to study the dynamic and
thermodynamic behaviour of such complex systems.

MD provides us with a microscopic, most often, classical~\footnote{Car 
and Parrinello~\cite{CP} have proposed a consistent quantum 
mechanical generalization of the MD approach.}
description of the system, which consists in following its time
evolution by solving numerically the Newton equations of motion of its
elementary constituents. 

The general mathematical setting of MD for the simulation of diffusive 
systems is well known~\cite{AT} and we will not dwell on it here.
For a general presentation of the method with a discussion of some of
the tricks that have been developed to efficiently implement MD codes
on the existing most powerful parallel platforms in the case of
($NVE$) simulations, we refer the reader to~\cite{NOI}.

We have prepared two chemically identical samples  of largely different 
size of our bilayer model, consisting of $2\times 32$ and $2\times
256$ molecules, hereafter denoted for short $BM(S)$ and $BM(L)$
(``$S$" and ``$L$" stand for ``small" and ``large", respectively). Since,
as we said in the Introduction, we do not have explicitly water in our
simulations, we will neglect electrostatic interactions altogether. For
consistency we have taken, as a model for the constituent
molecules, an artificial variant of DMPC, christened by us DMMG ({\it
Dimyristoyl-methyl-glycerol}), in which the charged part of the DMPC 
head (defined as the part of the molecule above the first atom--a P
atom--bound to the glycerol oxygen) has been replaced by a single CH$_3$
group. The detailed chemical structure of the Y-shaped DMPC molecule is
shown in Fig.~\ref{fig:schema1}, where we have also indicated the 
modification we have made on its structure in order to
construct the artificial variant of it we will be dealing with in this paper. 
%*************************************************************************** 
The elementary constituents of the DMMG molecule will be schematically
represented by LJ sites centered on the CH$_n$ groups ($n=1,2,3$),
thereby using the so-called ``united atom" approximation.  In this 
way each DMMG molecule will consist of 37 elementary units (for short simply
``atoms" in the following). Intra-molecular and fully flexible
inter-molecular interactions are described by making reference to the OPLS
force field~\cite{OPLS} with the modifications introduced in
ref.~\cite{WILSON}.  According to the general OPLS philosophy, 
we have reduced the strength of the
1-4 LJ potential in each dihedral angle by a factor 8.
The inter-molecular potential was
linearly switched off between 1.0 and 1.1~nm. As we already said, no electric
charges are attributed to atoms.

We have assumed the crystallographic density of our artificial system,
$\rho_{\circ}$, to be the same as that of DMPC~\cite{DENS}. Experimentally the
DMPC crystallographic surface density corresponds to an area of 0.396~nm$^2$
per molecule in the plane of the bilayer (the $x$--$y$ plane). Despite the
fact that we have drastically simplified the structure of the head of the
constituent molecules, we decided to keep the same number of molecules per
unit area as in the case of a DMPC bilayer, in order to be able to compare
the properties of our model with those of a somehow related real system. At
crystallographic density the surface of the system $BM(S)$ spans an area of
$3.56\times 3.56$~nm$^2$, that of the system $BM(L)$ an area 8 times larger,
equal to $14.24\times 7.12$~nm$^2$.

The separation of the two layers in the $z$ direction is fixed at the beginning
of each simulation by setting the distance between the branching points
(the C13 atoms in the picture of Fig.~\ref{fig:schema1}) 
of two opposite molecules, equal to 4.4~nm. 

In order to build the initial bilayer configuration we have started by
generating a molecule with the dihedral angles of molecule A in Table 1
of ref~\cite{DENS}. Tail~\#1 of this molecule~\footnote{To be definite
we call tail~\#1 the shortest of the two DMMG tails (the O14-C31
tail in Fig.~\ref{fig:schema1}) and tail~\#2 the longest one (the
C15-C46 tail).} was taken to lie in the $z$--$y$ plane, parallel to the
$z$ axis, with the head pointing in the positive $z$ direction. Notice
that tail~\#1 shows a kink in correspondence to the dihedral angle
O14-C17-C19-C20, making it parallel to tail~\#2 after C19. The basic
molecule of the lower layer is generated by performing a rigid rotation
of the first molecule by $180^\circ$ around the $y$ axis, holding fix
the C13 atom. The two C13 atoms are then displaced by 4.4~nm, by translating
one molecule with respect to the other one along the $z$ axis. The
fundamental cell of the initial crystal is an orthorhombic prism with
the square basis in the $x$--$y$ plane and the longer edge along the $z$
axis. Two copies of this pair of molecules have been placed in the cell with
the C13 atoms of the lowest molecules located one in a vertex and the other
in the center of the $x$--$y$ face, respectively. Finally the fundamental cell
is replicated in the $x$ and $y$ directions the necessary number of times
({\it i.e.}~4$\times$4 and 16$\times$8 times, respectively).

As usual, to limit finite-size effects, both $BM(S)$ and $BM(L)$
samples are taken to be periodic in the $x$ and $y$ directions. For
simplicity of computation, periodicity is also imposed in the $z$
direction, but adjacent copies of the bilayer are separated by about
100~nm, a distance so large that there are no possible interactions 
among different copies of the system.

Two different MD codes, expressly developed by our group, were 
employed for the simulations of the two systems. Both codes make use of an
MTS integration algorithm~\cite{MTS,PB} with a long time step of 5~fs. A short
time step 10 times smaller is introduced to integrate stretching and bending
contributions. The package used for the $BM(S)$ system ($2\times 32$
molecules) was run on a Digital 500 $\alpha$-station. The code produces 1~ps
of simulated dynamics every $\simeq$~100~s of CPU-time (corresponding to
0.5~s of CPU time for every long time-step for a system of 2368 atoms). For
the larger $BM(L)$ system ($2\times 256$ molecules) we have employed a
parallel version of the previous code. This code evolved from the one we
used in~\cite{NOI} to test the level of the performances offered by parallel
platforms in MD simulations of diffusive systems. Given the large request of
computing power necessary to perform simulations of such a big system, use
was made of the parallel platform called {\it Torre}, the largest (512 nodes,
25 Giga-flops) of the today available APE computers~\cite{APE}. On this
platform producing 1~ps of simulated dynamics costs $\simeq$~2000~s of
CPU-time (which corresponds to 2~s of CPU time for every long time-step for a
system of 18944 atoms).

In Tables~\ref{tab:tab1} and~\ref{tab:tab2} we report for the two samples
($BM(S)$ and $BM(L)$) we have constructed the set of values of surface
density and temperature ($T$ in Kelvin), at which simulations have been
carried out. The surface density is expressed in units of the
crystallographic density $\rho_{\circ}$ through the formula $\rho =
\mu\cdot\rho_{\circ}$. The numbers in the entries of the Tables
represent the length of the corresponding trajectory. 
%%%%%%%%%%%%%%%%%%%%%%%%%%%%%%%%%%%%%%%%%%%%%%%%%%%%%%%%%%%%%%%%
%%%%%%%%%%%%%  DEC  %%%%%%%%%%%%%%%%%%%%%%%%%%%%%%%%%%%%%%%%%%%%
%%%%%%%%%%%%%%%%%%%%%%%%%%%%%%%%%%%%%%%%%%%%%%%%%%%%%%%%%%%%%%%%
\begin{table}[!htbp] 
\begin{center}
\caption{The set of densities ($\rho=\mu \cdot \rho_{\circ}$,
$\rho_{\circ}=$ crystallographic density) and equilibration temperatures
($T$) at which we have run the simulations of the system $BM(S)$. The
numbers in the entries of the Table represent the length of each trajectory
in ns, not counting the initial velocity-rescaled steps.} 
\begin{tabular}{c||ccccccccc} \hline
 $T$(K)// $\mu$ & 1.05 & 1.0 & 0.95 & 0.91 & 0.87 & 0.83 & 0.79 &
0.76 & 0.69\\
\hline
150 & 0.9 & 5.4 & 0.9 & 0.9 & 0.9 & 5.4 & 0.9 & 0.9 & 5.4 \\
225 & 0.9 & 5.1 & 0.9 & 0.9 & 0.9 & 0.9 & 0.9 & 0.9 & 0.9 \\
250 & 0.9 & 5.1 & 0.9 & 0.9 & 0.9 & 0.9 & 0.9 & 0.9 & 0.9 \\
275 & 5.1 & 5.4 & 0.9 & 0.9 & 0.9 & 5.4 & 0.9 & 0.9 & 5.4 \\
300 & 0.9 & 5.1 & 0.9 & 0.9 & 0.9 & 0.9 & 0.9 & 0.9 & 0.9 \\
325 & 0.9 & 5.4 & 0.9 & 0.9 & 0.9 & 9.9 & 9.9 & 0.9 & 5.4 \\
350 & 0.9 & 5.1 & 0.9 & 0.9 & 0.9 & 5.4 & 0.9 & 0.9 & 0.9 \\
\hline
\label{tab:tab1}
\end{tabular}
\end{center}
\end{table} 

At every density value, the sample $BM(S)$ has been brought
at the required temperature by a simple velocity-rescaling algorithm,
which was run for more than 100 ps in the worst case. The surface
density of the bilayer was fixed at the chosen value at the
beginning of each simulation by uniformly rescaling the $x$ and $y$
coordinates of the center of mass of each molecule by the factor
$\sqrt{\mu}$. The initial configuration of the larger system, $BM(L)$,
was obtained by replicating the last available configuration of
$BM(S)$ with the desired values of temperature and density, and then
equilibrating the whole system for further 40~ps, again using a
velocity-rescaling algorithm. Throughout this paper we will indicate for
short by $T$ the equilibration temperature, {\it i.e.}~the temperature
held fixed during the velocity-rescaling steps, and by $T_s$ the real
temperature at which the simulation was actually run. Of course the two
temperatures will be somewhat different from each other. For most of our
considerations in this paper this difference will be of little
importance.

%%%%%%%%%%%%%%%%%%%%%%%%%%%%%%%%%%%%%%%%%%%%%%%%%%%%%%%%%%%%%%%%%%%%%
%%%%%%%%%%%%%%%%%%%%%%    APE     %%%%%%%%%%%%%%%%%%%%%%%%%%%%%%%%%%%
%%%%%%%%%%%%%%%%%%%%%%%%%%%%%%%%%%%%%%%%%%%%%%%%%%%%%%%%%%%%%%%%%%%%%
\begin{table}[!htbp]
\caption{Same as in Table~1 for the system $BM(L)$. Trajectory lengths
are in ps.}
\begin{tabular}{c||ccccccc}
\hline
 $T$(K)// $\mu$ & 1.05 & 1.0 & 0.95 &0.87 & 0.83 & 0.79 & 0.69 \\
\hline
150 & $\bullet$ & $140\;$ & $\bullet$ & $\bullet$ &$140\;$ &
$\bullet$& $140\;$ \\
225 & $40\;$ & $60\;$ & $40\;$& $40\;$ & $\bullet$ & $40\;$& $\bullet$ \\
275 & $40\;$ & $140\;$ & $60\;$& $40\;$ & $140\;$ & $40\;$ & $140\;$ \\
300 & $\bullet$ & $60\;$ & $\bullet$ & $\bullet$ & $\bullet$ & $\bullet$& 
$\bullet$ \\
325 & $40\;$ & $140\;$ & $40\;$ & $40\;$ & $140\;$ & $40\;$ & $140\;$ \\
350 & $40\;$ & $60\;$ & $40\;$ & $40\;$ & $\bullet$ & $40\;$ & $\bullet$ \\
\hline
\end{tabular}
\label{tab:tab2}  
\end{table}

Few comments are in order here.

1) The simulations of the small system $BM(S)$ are particularly long 
(they range from a minimum of 0.9~ns up to a maximum of 9.9~ns, not counting
the initial velocity-rescaling equilibration steps) and span a grid of 
$7\times 9 = 63$ points in the temperature-density plane
(Table~\ref{tab:tab1}). Notice that the trajectories employed in the
analysis presented in the next sections are all longer than 5~ns.

2) $BM(L)$ simulations cover a similarly large region of temperatures 
and densities with a total of 28 points. Trajectories are, however,
much shorter (Table~\ref{tab:tab2}).

\section{Simulation strategy}

One of the reasons to perform Molecular Dynamics simulations of such complex
systems, like cell membranes, is that many biological processes are still 
waiting for an explanation at molecular level. Not many of the experimental
techniques presently available may reach this goal at the resolution
required and under really well controlled physico-chemical conditions. On
the other hand, it is obvious that in order to be reasonably confident
about the validity of the model used in a simulation, the structural
properties of the model must be carefully compared to those of the ``real"
system so as to check the quality of the correspondence between
simulation results and actual experimental data.

\begin{figure}[!htbp]
\centering
\includegraphics[width=10cm]{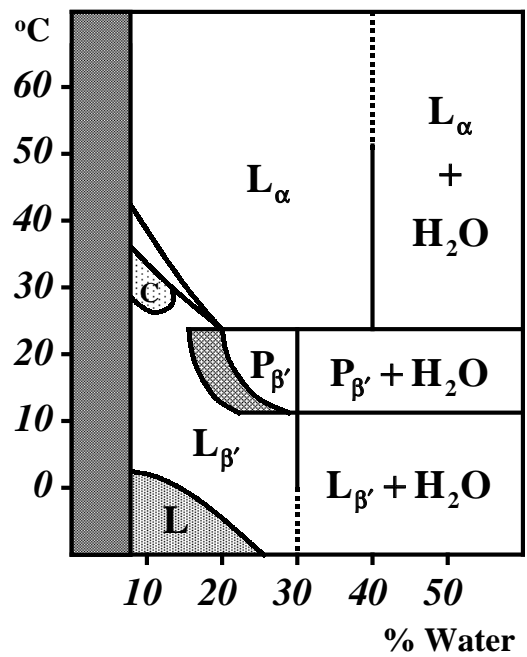}
\caption{The typical phase diagram of a phospho-lipid bilayer. The figure is 
taken from ref.~\cite{SSSC} and refers to a {\it Dimyristoyl-lecitin} 
bilayer.}
\protect\label{fig:phase}
\end{figure}

In the case we are investigating, one of the most interesting macroscopic 
features characterizing the biological behaviour of the system is 
the structure of its phase diagram. The typical phase diagram of a 
phospho-lipid bilayer is reproduced in Fig.~\ref{fig:phase}. 
%***************************************************************************
%***************************************************************************
The figure is 
taken from ref.~\cite{SSSC} and refers to a {\it Dimyristoyl-lecitin}
bilayer. As it is also seen from the figure, many phases have been identified
experimentally: the most clearly separated ones are those called L, L$_\beta$
and L$_\alpha$, corresponding to the crystal, gel and liquid-crystal
phase of the system, respectively. It is important to note here for the future
comparison with simulation data that the horizontal axis in
Fig.~\ref{fig:phase} is the percentage of water per phospho-lipid molecule
and not directly the surface density. As we have argued before, however, to
first approximation these two quantities are proportional to each other, as
an increase of water concentration has mainly the effect of ``diluting" the
bi-dimensional fluid formed by the molecular heads. 

The nature of the structural changes associated to the different phase
transitions has been mainly investigated in NMR experiments. NMR
data~\cite{NMR} indicate that, starting from the ordered crystal
phase, the transition to the gel phase corresponds to an order-disorder
transition of the heads, in which the tails retain their order, but undergo
a collective ``tilt" with respect to the plane of the bilayer~\cite{SSSC}.
The transition to the liquid-crystal phase is related to a further
order-disorder transition, which this time involves also the tails.

The experiments that put in evidence the existence of these different phases
are generally performed on liposomes (sort of artificial small cells) in
solution or on the so-called Langmuir-Blodgett multi-layers and are commonly
performed at constant pressure. In these conditions the surface density 
of phospho-lipid heads changes with temperature. On the simulation side the
requirement of holding the pressure constant can be fulfilled by
performing MD simulations in the ($NpT$) {\it ensemble}. Simulations of
this type are rather delicate as they require sophisticated strategies to
force the system to move in the phase space with the appropriate
statistical weight and to tame otherwise annoying numerical instabilities.
It should be immediately said, however, that today a lot of skill has been
developed to properly deal with all these
problems~\cite{K,KLEIN_DPPC,ZXMT}. Simulations in the ($NVE$) ({\it
micro-canonical}) {\it ensemble} are definitely much simpler and numerical
stability is easily achieved. In the {\it{micro-canonical ensemble}}
pressure and temperature are not independent variables and can in principle
be determined
once the values of the density, $N/V$, and total energy, $E$, are given. 

Our strategy was to make a systematic study of the structural properties
of the system by performing MD simulations at many different values of the
($NVE$) parameters in such a way as to explore quite a large set of values
of temperatures and densities. The high ``resolution" of our exploration has
allowed us to draw interesting conclusions about the properties of the
crystal and gel phase of the system. Results obtained for $BM(S)$ are
confirmed by the analysis of the data we collected for the much larger
system, $BM(L)$. As we shall see, $BM(L)$ data are affected by almost
identical statistical errors as the $BM(S)$ data, though the available
trajectories are much shorter.

\subsection{Extracting physical information from simulations}

Generally speaking, in order to show the existence of a phase transition one
must be able to define an appropriate set of ``order parameters" which
could be used to describe the structure of the phase diagram of the
system. The degree of order/disorder of the system can then be monitored
through functions and parameters characterizing the translational and/or
the orientational distribution functions of the constituent molecules. As a
general reference for this kind of analysis, we shall follow in this paper 
the approach used in the discussion of liquid-crystal simulations in 
ref.~\cite{LQ}.

As a first step in the study of the order properties of the system, we 
define the average direction of the constituent molecules. This quantity is
represented by the 1-st rank order parameter  
\begin{equation} 
\vec{P}^{(\ell)}=\langle \langle \vec{u} \rangle \rangle^{(\ell)} 
\label{P1}
\end{equation}  
where, for elongate molecules, like the ones we are dealing with, the unit
vector $\vec{u}$ is naturally taken as the principal axis of the inertial
tensor of the molecule ({\it i.e.}~the normalized eigenvector of the
inertial tensor with the lowest eigenvalue). The double bracket in
eq.~(\ref{P1}) means average over both the $N_{mol}$ (identical) molecules
of the system and over the $N_{conf}$ configurations collected in the
simulation. The superscript ``$\ell$" ($\ell$=1,2) means that the average is 
restricted to the $\ell$-th layer (conventionally we decided to call
layer~\#1 the upper layer and layer~\#2 the lower one).
The $z$ component of $\vec{P}$, which is usually called $P_1$,
will be used to monitor the average orientation of the molecules. 

A further order parameter, which is also experimentally accessible
(mainly in $^2$H-NMR experiments~\cite{NMR}), that can be used to parametrize 
the local orientational order of molecular segments, is   
\begin{equation}  
S^{(\ell)}_i=\langle \langle \frac{1}{2}(3\cos^2 \beta_i -1)\rangle
\rangle^{(\ell)} \label{OP} 
\end{equation}   
where the superscript ``$\ell$'' has the same meaning as in eq.~(\ref{P1}). 
In NMR
experiments $\beta_i$ is the  angle between the $i$-th carbon-deuterium
(C$_i$-D) bond in a tail and the external static magnetic field (which is
usually taken to coincide with the normal to the bilayer surface). $S_i$,
often called ``segmental order parameter", defines a good order parameter as
it enjoys the following characteristic properties          
\begin{itemize} 
\item When the C$_i$-D bonds are essentially aligned in a given direction 
with respect to the magnetic field, the system is in an ordered phase with
$S_i \simeq \frac{1}{2}(3\cos^2 \beta^{(0)} -1)\neq 0$, where $\beta^{(0)}$
is the most represented angular value in the distribution. In particular, if
all the bonds have $\beta_i=0$, then $S_i=1$.    
\item When the distribution of $\cos \beta_i$ is flat ({\it i.e.}~when all
the values of $\cos \beta_i$ appear with equal probability), $S_i
= 0$ and the system is in a completely disordered phase.    
\end{itemize}
 
The crucial point that must be taken into account in trying to adapt the 
above definition to the situation one encounters in numerical simulations
is that the order parameter we would like to define should allow us to
distinguish in a clear way genuine order--disorder transitions from
possibly present (and also very interesting) transitions between different
kinds of ordered structures. The latter is precisely the situation that
occurs experimentally in the crystal$\rightleftharpoons$gel transition,
where, as we recalled above, the phase transition appears to be
characterized by the following two structural changes: 1) heads undergo an
order--disorder transition, 2) tails remain ordered, but the angle made by
their average direction with the normal to the plane of the bilayer
increases significantly (tilt). The problem with the definition~(\ref{OP})
is that, in the situation in which the reference axis is taken to be the
same for all tail segments and lie parallel to the bilayer normal, an
overall change in the direction of the tails with respect to it will
actually affect the value of the order parameter (making it smaller), much
in the same way a real increase of the disorder of the tails would do.

To overcome this type of difficulties it was suggested in~\cite{LQ}
that the way to appropriately monitor the local orientational order
of atomic segments along the molecule is to use the formal expression
given in eq.~(\ref{OP}), but with $\beta_i$ the Euler angles formed by the
$z$ axes of suitably chosen local reference frames (see below) with
the average direction of the molecules in the layer to which the segments
one is considering belong. Each local frame is defined by taking the $z$
axis parallel to the segment connecting two non-contiguous carbon atoms
along a tail.

Let us now see how these general considerations are translated into 
precise formulae. Since we are interested in investigating 2-nd rank
tensor properties of the system, as those that are extracted from NMR
experiments, it is convenient to measure order with respect to the
directions of ``maximum 2-nd rank order"~\cite{AT,ZAN}.
These directions are determined starting from the construction of the
traceless tensor (whose definition closely parallels that of quadrupole
moment of a charge distribution~\cite{SLI})  
\begin{equation}  
Q^{(\ell)}_{\alpha,\beta}=\langle\langle \frac{1}{2}
(3~u_{\alpha}~u_{\beta} - \delta_{\alpha\beta}) \rangle\rangle^{(\ell)}
\qquad \alpha,\beta=1,2,3 
\label{Q}
\end{equation}   
where $\vec{u}$ is a unit vector related to the geometry of the molecules,
$\delta$ is the Kronecker symbol and again one defines a $Q$-tensor for each
one of the two layers. Eigenvalues and eigenvectors of $Q$ reflect
collective orientational properties of the system. For instance, if the
phase is  uniaxial (more mathematically, if the system is a collection of
objects with $D_{\infty,h}$ point space group symmetry), $\vec{u}$ should
be taken as the direction of the principal axis of inertia of the
molecules. Then $Q$ will have one positive eigenvalue, say, $\lambda$
(usually called $P_2$) and two identical negative eigenvalues,
$-\lambda/2$. A biaxial phase is characterized by the fact that the two
negative eigenvalues are not equal, while a more complicated orientational
order is reflected in a non-trivial dependence of the eigenvectors of $Q$
on the choice of $\vec{u}$. In the uniaxial phase the (normalized)
eigenvector corresponding to the highest eigenvalue, $\lambda$, is called
``director" and will be indicated by $\vec{d}^{(\ell)}$ in the following.
In our case $\vec{d}^{(\ell)}$ represents the average direction of
alignment of the long molecular axes in layer $\ell$.

A precise notion of local order in each layer is introduced by comparing 
the orientation of a local rigid frame associated to any given molecular
fragment with the orientation of a fixed reference frame, related to the
collective order of the system, which we will identify with the
(orthogonal) eigenvectors of $Q^{(\ell)}$. Given three non-aligned atoms,
indexed by $i_1$, $i_2$ and $i_3$ the local frame associated to each fragment,
$i$ $(i\equiv i_1 i_2 i_3)$, 
is constructed by taking the $z$ axis as the vector
joining the atom $i_1$ with the atom $i_3$ and the $y$--$z$ plane as the
plane where the three atoms lie. The azimuthal and polar angles ($\alpha_i$
and $\beta_i$) of the director $\vec{d}^{(\ell)}$ in these local frames can
be used in turn to specify their orientation.

It is important to remark that, in comparing values of the order parameter,
$S_i$, constructed through eq.~(\ref{OP}) where the angles $\beta_i$ are
defined as described above, with numbers extracted from NMR experiments,
the latter must be multiplied by a suitable numerical factor. This is
related to the fact that in our definition of order parameter we monitor the
orientation of the segment joining, say, the two non-contiguous
C$_i$-C$_{i+2}$ atoms, while the two experimentally relevant C$_{i+1}$-D
bonds, whose average direction is extracted from NMR data, lie in a plane
almost perpendicular to it. Given the relative orientation of the different
reference systems we have introduced, it can be shown, using Wigner
rotation matrices, that in our geometrical conditions this factor is always
very near to $-2$~\cite{NMR}. 

To monitor the translational order/disorder properties it is customary to 
make reference to the pair distribution function. Given the strong
anisotropy of the molecule arrangement, typical of our system, it is
convenient to define the two pair distribution functions,
$g_{\parallel}(r_\parallel)$ and $g_{\perp}(r_{\perp})$, where
$r_{\parallel}$ and $r_{\perp}$ represent the length of the components of
the vector joining the atoms of the pair in the direction parallel to the
bilayer normal and in the plane perpendicular to it, respectively. As
usual, $g_{\parallel}$ and $g_{\perp}$ are normalized to the corresponding
ideal gas distributions at the same density. We thus write       
\begin{equation}   
g_{\parallel} = \frac{h_{\parallel}}{h^g_{\parallel}}
\qquad g_{\perp} = \frac{h_{\perp}}{h^g_{\perp}}   
\label{gperp}
\end{equation}
where $h(r)$ is the number of pairs of atoms at (parallel, or respectively
perpendicular) distance equal to $r$ and $h^g(r)$ is the corresponding
quantity for the ideal gas at the same density. It should be noted that we
separately define a $g_{\perp}$ distribution for each one of the two
layers by correspondingly limiting the calculation to the appropriate set of
molecules.  

\subsection{Data analysis}

For an ergodic system {\it micro-canonical} averages are computed as time
averages of infinitely long trajectories. Such time averages are approximated 
in practice by averages over the largest possible number of independent
configurations one can collect (given the available computing power)
and representative of the infinite set of configurations of the system. 
As for the statistical error to be attributed to these {\it ensemble}
averages, one should take the associated standard deviation.

A look at the time evolution of the $BM(S)$ system and, in particular,
at the behaviour of the inter-molecular energy ($E_{inter}$) shows that a
plateau is smoothly reached in most cases only after about 2.5-3~ns. In view
of this rather slow equilibration time we prudentially decided to exclude
from the analysis presented in the next section the first 3~ns of each
MD trajectory.  

The situation is much better for the $BM(L)$ system for two reasons. One is
that, as we explained before, the initial configuration of each $BM(L)$
simulation was obtained by replicating the (already well equilibrated) last
available configuration of the $BM(S)$ system and further proceeding with  
an equilibration of the full system. The second is that the equilibration
time of a larger system is naturally shorter.

To decide how distant in time two successive configurations should be in
order to consider them as uncorrelated, we studied the auto-correlation
function of the potential energy. In the case of the $BM(S)$ system, we
found that, after removing from the analysis the configurations referring
to the first 3~ns of each trajectory for the reasons explained above, the
auto-correlation function dies away in a few ps at all values of density and
temperature. A similar situation occurs in the case of the $BM(L)$ system
(though all the available configurations, produced after the initial
velocity-rescaling equilibration, were kept in the analysis). Accordingly,
we then decided to record a configuration every 1000 long time-step
iterations ({\it i.e.}~every 5~ps) in all cases.

For the small system this means that we will have at our disposal not less
than 480 uncorrelated configurations (and sometimes much more, up to
1380, see Table~\ref{tab:tab1}), that can be used in the forthcoming
analysis. For the larger system $BM(L)$, where much shorter trajectories
have been collected, averages will be taken over a significantly smaller
number of configurations. Despite this fact, thanks to the much larger size
of the sample $BM(L)$ compared to that of $BM(S)$, 
statistical errors turn out to be of comparable magnitude.

\section{Results}

Out of the many surface densities we have explored we will show data
corresponding to the three cases, $\mu$ = 1.0, 0.83 and 0.69, that roughly
represent the experimental densities of the three most significant phases 
of the DMPC bilayer, {\it i.e.}~crystal, gel and liquid-crystal,
respectively. For each of these densities we will discuss results at three
different temperatures (Tables~\ref{tab:tab1} and~\ref{tab:tab2}). The
temperatures we have considered were chosen so as to have them below, 
in between and above the two most significant phase transitions 
(crystal$\rightleftharpoons$gel and gel$\rightleftharpoons$liquid-crystal)
the DMPC bilayer undergoes (see Fig.~\ref{fig:phase}). 

Despite the fact that we do not regard the results we
obtained at the lowest density as physically interesting for the description
of the expected liquid-crystal phase of a bilayer, we will nevertheless
briefly discuss them in this section, mainly to contrast these data with what
we get at the more reliable values of the density, $\mu$=1.0 and $\mu$=0.83.
This discussion may also be very useful to identify where and how 
our model should be improved. 
%\vskip 6cm
%***************************************************************************
\begin{table}i[!htbp] 
\caption{Inter-molecular ($E_{inter}$) and intra-molecular ($E_{intra}$)
energies at some selected temperature-density points for the $BM(S)$ system. 
Numbers in the column labeled by $T_s$ are the measured temperatures
at which the corresponding simulation has actually run. Errors are in 
brackets.}  
\begin{tabular}{cc||cccccccc}
\hline
$T$ (K) & $\mu$ & $T_s$ (K) & 
 $E_{inter}$ (kJ/mol) & $E_{intra}$ (kJ/mol)\\
\hline
150 & 1.00 & 157 (2) & -207.5 (0.6) & 84.8 (1.0) \\
150 & 0.83 & 151 (2) & -201.0 (0.6) & 85.6 (0.9) \\
150 & 0.69 & 160 (2) & -195.9 (0.6) & 96.5 (1.0) \\
\hline
275 & 1.00 & 283 (4) & -204.1 (1.0) & 131.7 (1.6) \\
275 & 0.83 & 281 (3) & -196.7 (1.0) & 132.3 (1.6) \\
275 & 0.69 & 294 (4) & -186.4 (1.0) & 143.9 (1.6) \\
\hline
325 & 1.00 & 338 (4) & -199.7 (1.2) & 154.6 (1.9) \\
325 & 0.83 & 341 (4) & -186.7 (1.2) & 161.0 (1.9) \\
325 & 0.69 & 343 (5) & -185.0 (1.7) & 175.8 (2.1) \\
\hline
\end{tabular}
\protect\label{tab:tab4}   
\end{table} 

\subsection{The system $BM(S)$}

We shall start our analysis by discussing the data we get from the
simulations of the system $BM(S)$, for which we have collected very long
trajectories (see Table~\ref{tab:tab1}). We will use the results coming from
the $BM(L)$ system to refine and confirm the physical picture that emerges.
   
\subsubsection{Thermodynamic properties}
%\vskip .2cm  
%\noindent{\it \underline{Thermodynamic properties}}  
%\vskip .2cm 
The density and temperature dependence of $E_{inter}$ and $P_2$ should show 
marked variations in correspondence to first or second order phase 
transitions. In Table~\ref{tab:tab4} a smooth and rather flat dependence of
$E_{inter}$ on $T$ and $\mu$ is, instead, seen~\footnote{In the third
column of Table~\ref{tab:tab4} we have reported the actual physical
temperature, $T_s$, of each simulation.}. This may be partly due to a still
too coarse grid of points in the temperature and density plane and partly
to the finite size of the simulated systems. 
%**************************************************************************
\begin{table}[!htbp] 
\caption{Values of the $1^{st}$ rank order 
parameter, $P_1$, at some selected
temperature-density points, separately for the upper (\#1) and
the lower (\#2) layer of the $BM(S)$ system. Errors are in brackets.} 
\begin{tabular}{cc||cc} 
\hline
 $T$ (K) & $\mu$ & 
 $P_1$(layer~$\#1$) & $P_1$(layer~$\#2$) \\ 
\hline
150 & 1.00 & -0.909 (0.004) & 0.891 (0.004) \\
150 & 0.83 & -0.775 (0.003) & 0.763 (0.004) \\
150 & 0.69 & -0.637 (0.005) & 0.627 (0.004) \\
\hline
275 & 1.00 & -0.923 (0.005) & 0.913 (0.005) \\
275 & 0.83 & -0.788 (0.005) & 0.782 (0.005) \\
275 & 0.69 & -0.63 (0.02) & 0.58 (0.02) \\
\hline
325 & 1.00 & -0.914 (0.007) & 0.924 (0.005) \\
325 & 0.83 & -0.795 (0.007) & 0.800 (0.007) \\
325 & 0.69 & -0.46 (0.03) & 0.73 (0.02) \\
\hline
\end{tabular}
\label{tab:tab5a}
\end{table}

At all values of density and temperature the molar heat capacity per atom,
$c_V$, which was computed according to~\cite{LEB,NOI}, shows values
in the range 3--3.5 (in units of the gas constant, $R$), which are typical
of a molecular liquid.

As for the parameter $P_1$ (eq.~(\ref{P1})), we see from
Table~\ref{tab:tab5a} that it is rather sensitive to the average orientation
of the molecules with respect to the $z$-axis (which, we remember, coincides
with the bilayer normal) and, as expected, it steadily decreases with the
density. 
\begin{table}[!htbp]
\caption{Same as in Table~\ref{tab:tab5a} for the $2^{nd}$
rank order parameter, $P_2$.} 
\begin{tabular}{cc||cc} \hline
 $T$ (K) & $\mu$ & 
$P_2$(layer~$\#1$) & $P_2$(layer~$\#2$) \\ 
\hline
150 & 1.00 &  0.986 (0.002)& 0.996 (0.001)  \\
150 & 0.83 &  0.979 (0.004) & 0.962 (0.003) \\
150 & 0.69  &  0.961 (0.004)& 0.949 (0.005) \\
\hline
275 & 1.00 &  0.996 (0.001) & 0.995 (0.001) \\
275 & 0.83 &  0.995 (0.001) & 0.995 (0.001) \\
275 & 0.69 &  0.95 (0.02) & 0.80 (0.02) \\
\hline
325 & 1.00 &  0.996 (0.001) & 0.995 (0.001) \\
325 & 0.83 &  0.978 (0.008) & 0.990 (0.002) \\
325 & 0.69 &  0.28 (0.05) & 0.84 (0.02) \\
\hline
\end{tabular}
\label{tab:tab5b}
\end{table}

We remark that
the values of the parameters $P_1$, $P_2$ and $d_z$ (see 
Tables~\ref{tab:tab5a}, \ref{tab:tab5b} and \ref{tab:tab5c}) 
for the two layers are consistent with each other within three (most often, 
two) standard deviations for the two highest values of $\mu$ ({\it
i.e.}~$\mu$=1.0 and 0.83). There is instead a statistically significant
difference, especially at the highest temperatures, if one compares values
referring to the upper and lower layer at the lowest density. This is due to
rare events in which some of the molecules in the upper/lower layer rotate
upward/downward relatively to the bilayer plane. Unfortunately the
characteristic times of these motions are too long compared to the length of
the trajectories we have produced to be in position of adequately sampling this
dynamics. It is then not surprising to find that, within the time window of
our simulations, the two layers do not appear statistically identical.
\begin{table}[!htbp]
\caption{Same as in Table~\ref{tab:tab5a} for
the director $z$ component, $d_z$.} 
\begin{tabular}{cc||cc} \hline
 $T$ (K) & $\mu$ &
 $d_z$(layer~$\#1$) & $d_z$(layer~$\#2$) \\
\hline
150 & 1.00 & 0.913 (0.004) & 0.892 (0.004) \\
150 & 0.83 & 0.781 (0.003) & 0.773 (0.004) \\
150 & 0.69 & 0.646 (0.005) & 0.639 (0.004) \\
\hline
275 & 1.00 & 0.924 (0.005) & 0.914 (0.005) \\
275 & 0.83 & 0.790 (0.005) & 0.783 (0.005) \\
275 & 0.69 & 0.67 (0.02) & 0.1 (0.7) \\
\hline
325 & 1.00 & 0.915 (0.007) & 0.926 (0.006)  \\
325 & 0.83 &  0.801 (0.007) & 0.803 (0.007) \\
325 & 0.69 &  0.6 (0.2) & 0.83 (0.01) \\
\hline
\end{tabular}
\label{tab:tab5c}    
\end{table} 

Although we do not see any sign of instability of our bilayer model 
at anyone of
the values of density and temperature we have explored (we have performed
simulations at densities as low as $\mu$=0.69 and temperatures as high as
$T$=350~K) or after very long simulation times (some of the trajectories are as
long as $\sim$10~ns), we will not consider data referring to $\mu$=0.69 as
physically significant, in view of the undesired behaviour described above.

All the data presented in this section are
consistent with the  description of the bilayer as an 
oriented molecular liquid,
if one excludes the density $\mu$=0.69, where the system starts to show an
increasing degree of isotropy. As for other structural parameters,
particularly the pair distribution functions (see below), they show a high
level of translational order that is progressively lost when the density
decreases or the temperature increases. Finally, it is worth noticing that
decreasing the density affects the orientational order in a stronger way
than increasing the temperature, indirectly suggesting that ordering is more
sensitive to hydration than to heating.   
\subsubsection{Structural properties}
\begin{figure}[!htbp] 
\centering
\includegraphics[width=10cm]{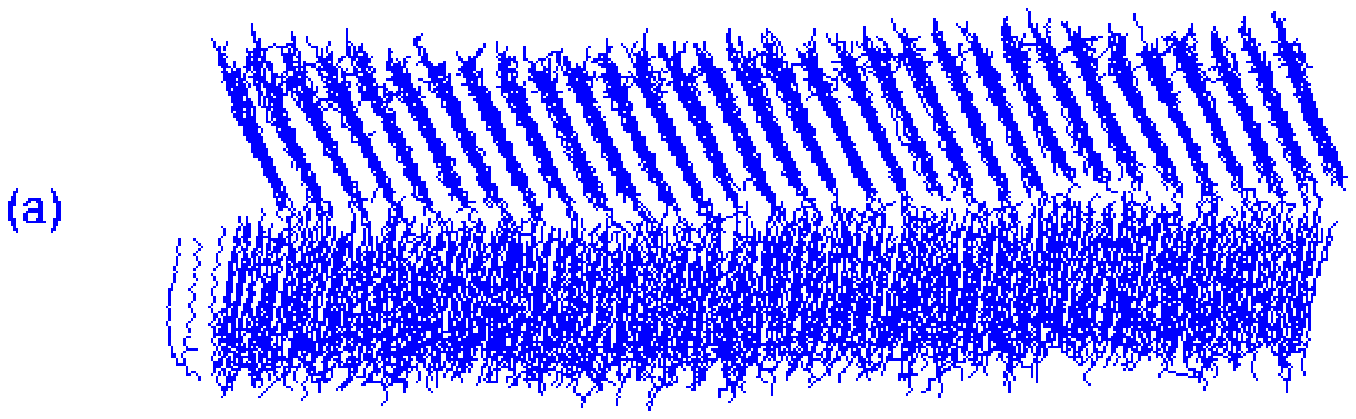}
\includegraphics[width=10cm]{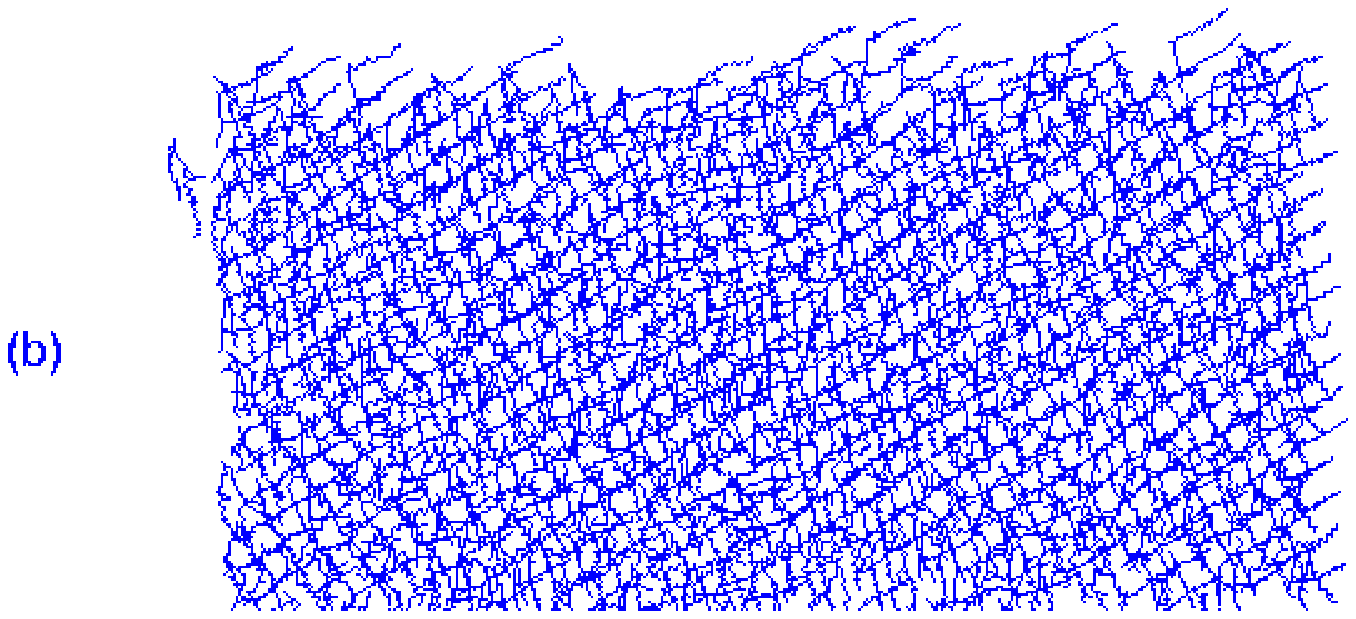}
\vskip 2cm
\caption {The final configuration of the $BM(L)$ simulation at 
$\mu$=1 and $T$=325~K: side view (a), top view (b).}
\protect\label{fig:crystal}
\end{figure}

\begin{figure}[!htbp]
\centering
\includegraphics[width=10cm]{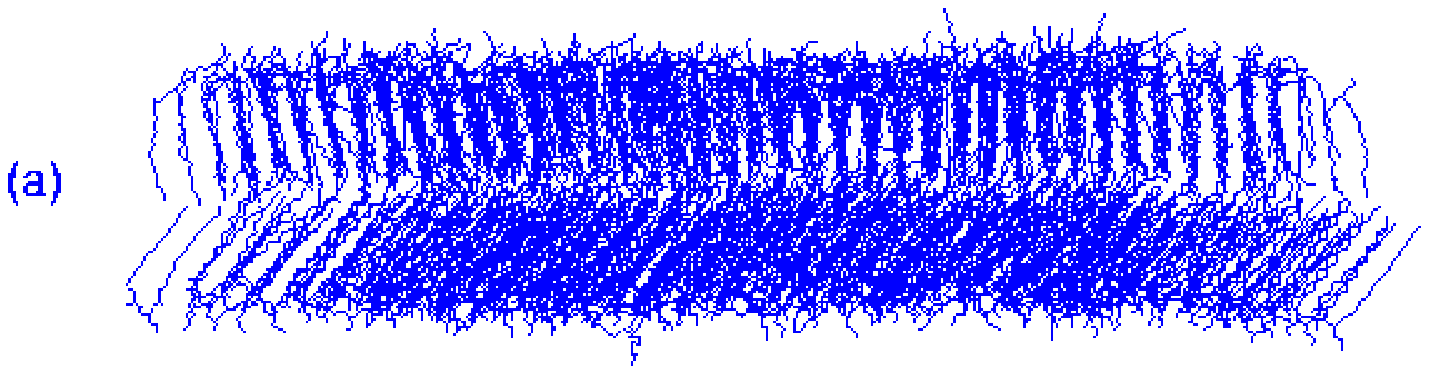}
\includegraphics[width=10cm]{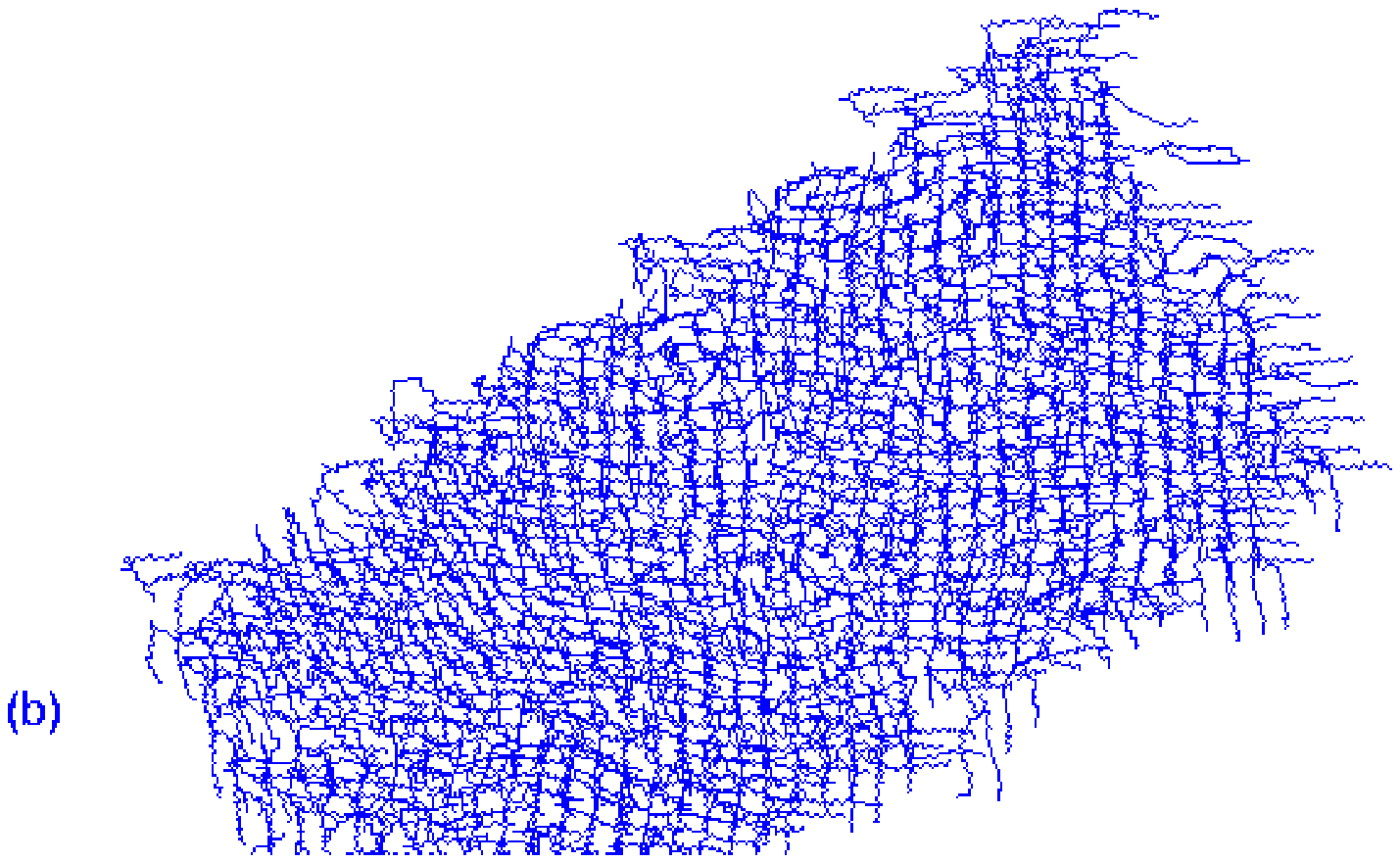}
\vskip 2cm
\caption{Same as in Fig.~\ref{fig:crystal} at $\mu$=0.83 and $T$=325~K.}
\protect\label{fig:gel}
\end{figure}

%\vskip .2cm
%\noindent{\it \underline{Structural properties}}
%\vskip .2cm
We start by discussing some general structural properties of the bilayer at
selected density-temperature points, that are suggested by a simple graphical
inspection of the geometry of the system. In Figs.~\ref{fig:crystal}
and~\ref{fig:gel} we show two views ((a) and (b)) of two snapshots of the
larger bilayer, $BM(L)$, at $T$=325~K and at the two densities, $\mu$=1 and
0.83, respectively. The snapshots are obtained using the program
MOLMOL~\cite{MOLMOL}. In panel (a) of Figs.~\ref{fig:crystal}
and~\ref{fig:gel} we show the side view of the bilayer as seen from the
direction in the $x$--$y$ plane from which the translational order of the
upper layer is most clearly visible. In panel (b) of Figs.~\ref{fig:crystal} 
and~\ref{fig:gel} we show the top view of the same pictures, obtained after
rotating the bilayer by 90$^{\circ}$ around an axis orthogonal to both the $z$
axis and the direction of ``best view" identified above. As a whole, the
figures show that the main effect of decreasing the density is a progressive
tilt in the average orientation of the molecules with respect to the bilayer
normal.
%***************************************************************************
%***************************************************************************
%***************************************************************************
%***************************************************************************
%***************************************************************************

\begin{figure}[!htbp]
\centering 
\includegraphics[width=10cm]{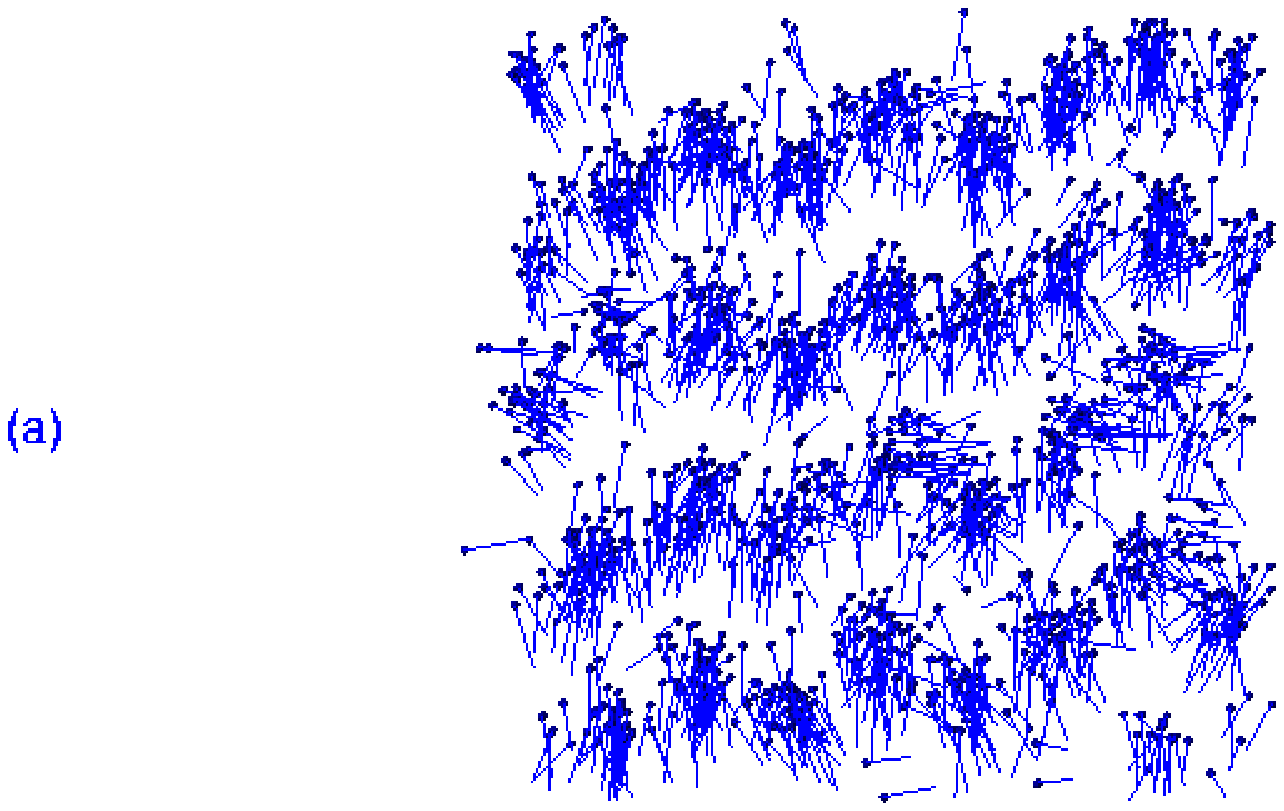} 
\includegraphics[width=10cm]{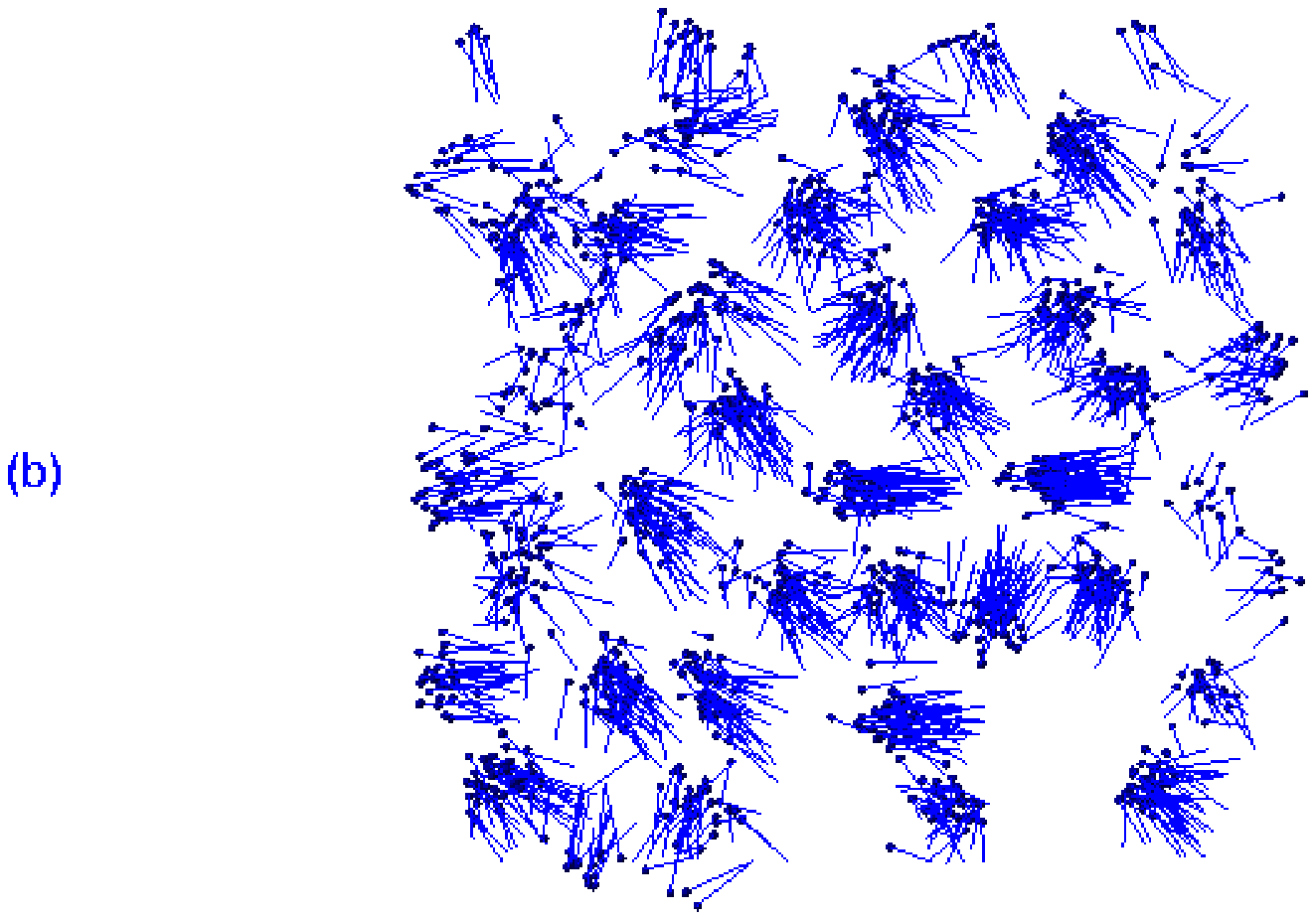}
\vskip 1cm
\caption{The projections on the $x$--$y$ plane of the vectors joining the 
pair of atoms C10-C13, as seen in selected configurations of the $BM(S)$
trajectory at $T$=325~K and $\mu$=1 (a), $\mu$=0.83 (b).}
\protect\label{fig:xyproj}
\end{figure}

\begin{figure}[!htbp]
\centering 
\includegraphics[width=10cm]{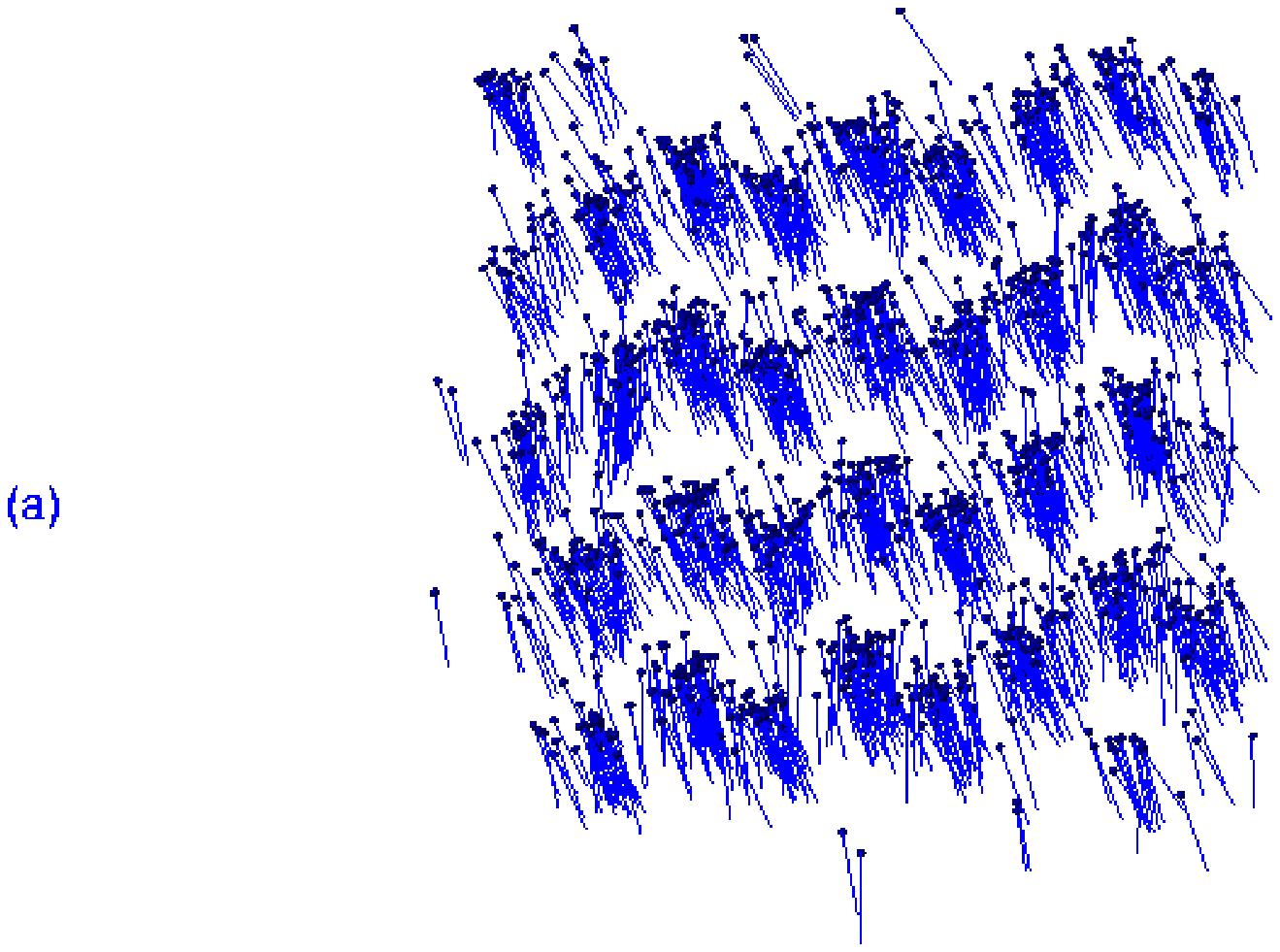} 
\includegraphics[width=10cm]{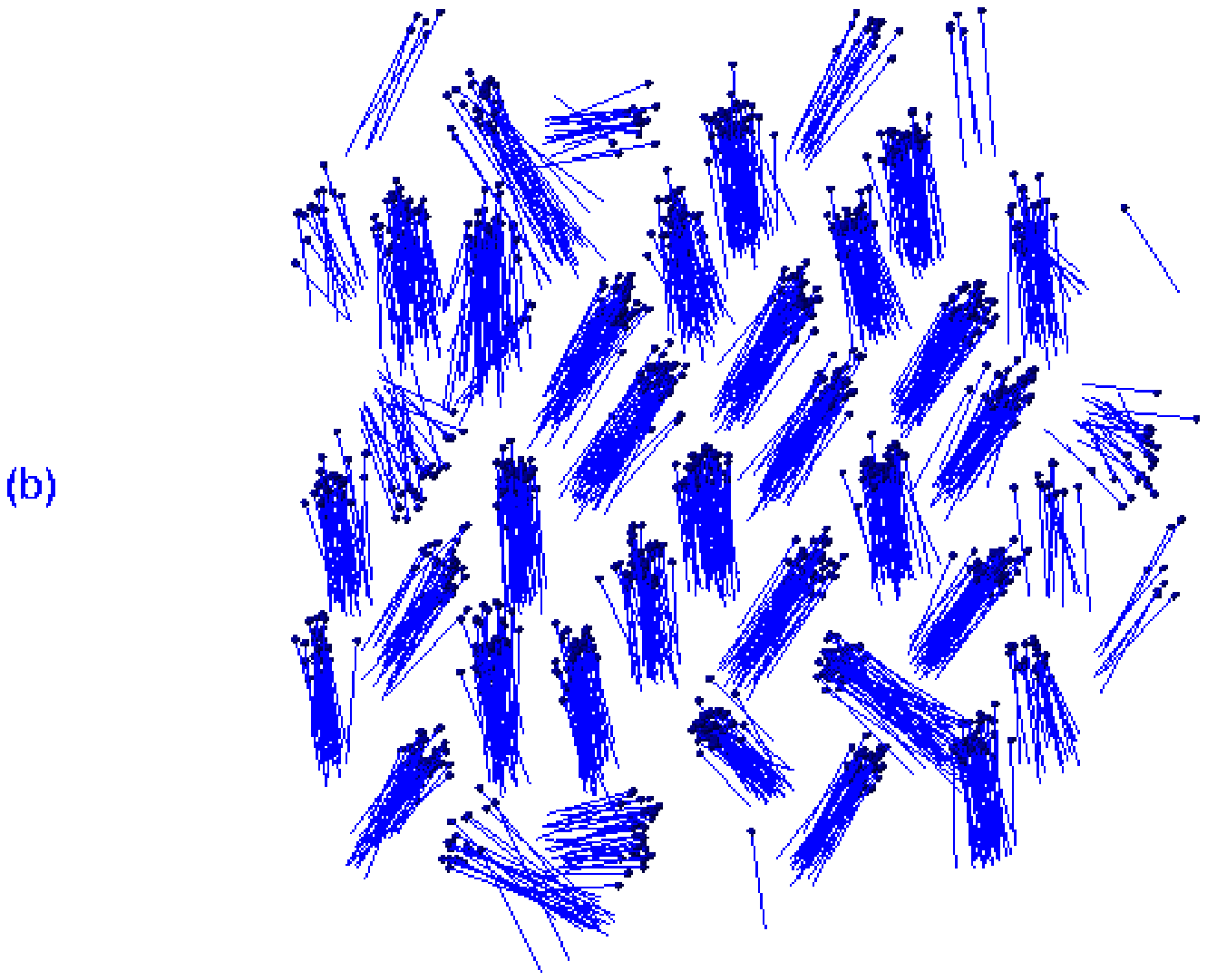}
\vskip 1cm
\caption{Same as in Fig.~\ref{fig:xyproj} for the C24-C37 pair of atoms.}
\protect\label{fig:xyproj2}
\end{figure}

Further information on the translational order properties of the system can 
be gained by looking at Figs.~\ref{fig:xyproj} and~\ref{fig:xyproj2}, where
we have plotted at the two (density, temperature) values considered above,
namely (a)=(1, 325~K) and (b)=(0.83, 325~K)), the projections on the 
$x$--$y$ plane of the segment joining the two atoms of the pair C10-C13
(Fig.~\ref{fig:xyproj}) and C24-C37 (Fig.~\ref{fig:xyproj2}), as they evolve
in time. 
%***************************************************************************
%***************************************************************************
%***************************************************************************
%***************************************************************************
The two pairs we consider are taken to belong to the upper layer. They differ
in their average distance from the middle plane lying in between the two
layers: the C10-C13 pair can be used to describe the orientation of the head,
as the two atoms belong to the top portion of the lipid molecule, while the
C24-C37 pair can be used to monitor the orientation of the imaginary
rectangle roughly containing the two molecular tails, because the C24 and C37
atoms belong to different tails and are located at more or less the same
height from the C13 joint. A remarkable feature, quite evident in all the
figures, is that molecules undergo fluctuations in local positional energy
minima that are spatially well ordered at high density and progressively less
ordered as the density decreases.
%***************************************************************************
%***************************************************************************
An other interesting question that can be studied by analyzing 
Figs.~\ref{fig:xyproj} and~\ref{fig:xyproj2} is the effect of density on the
degree of molecule packing: the range of positional fluctuations at $\mu$=0.83
is smaller than that at the higher density, $\mu$=1, both for heads
(Fig.~\ref{fig:xyproj}) and tails (Fig.~\ref{fig:xyproj2}). This means that
the larger space the molecules have at their disposal, going from higher to
lower densities, does not directly influence their actual translational
freedom. The reason for this behaviour may be due to the fact that a decrease
of the interaction potential among the molecules in each layer (that occurs
because molecules are more separated) is at least partly compensated by the
stronger interaction potential between the two layers (which follows from the
fact that an increase of the molecular tilt angle decreases the distance
between the two layers).

Finally, an inspection to the geometry of single molecules in different
density-temperature conditions shows that at $\mu$=1, $T$=150~K, the longer
C15-C46 tail (tail~\#2) lies essentially parallel to the head, while the
O14-C31 tail (tail~\#1) shows a kink in correspondence to the dihedral angle
O14-C17-C19-C20, making it parallel to tail~\#2 after C19. This geometric
arrangement, that is present in the initial  crystallographic
structure~\cite{DENS}, becomes ``inverted" at smaller densities and even at
$\mu$=1 at the highest temperature, $T$=325~K, in the sense that the kink
appears instead in the longer tail~\#2. In this situation  the two tails
happen to have almost the same effective length, if counted from the common
branching point, C13. The related torsional changes occurring in the
neighborhood of the branching region seems to be a very important ingredient
in arranging the internal molecular structure so as to comply with the
tilting and packing of molecules taking place in each layer.  

\subsubsection{Pair distribution functions}
%\vskip .2cm
%\noindent{\it \underline{Pair distribution functions}}
%\vskip .2cm
The information obtained through the graphical investigations we have just
described can be quantified in a useful way with the help of the pair
distribution functions and the order parameters defined in the previous
section.  
%***************************************************************************
\begin{figure}[!htbp]
\centering
\includegraphics[width=10cm]{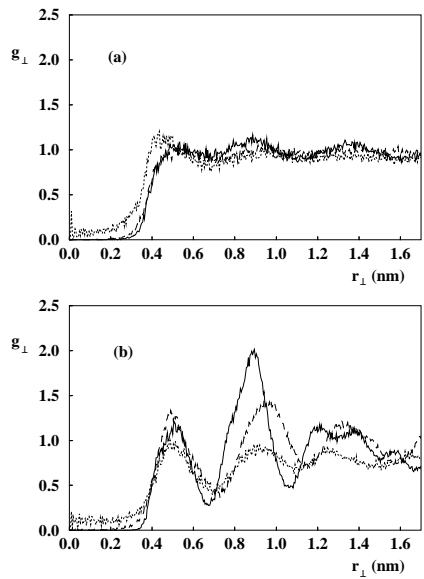}
\caption{The $g_{\perp}$ pair distribution function of C10 (a)
and C24 (b) atoms for the system $BM(S)$ at $T$=325~K and $\mu$=1 (solid
line), $\mu$=0.83 (dashed line), $\mu$=0.69 (dotted line).}
\protect\label{fig:gperpdgl}
\end{figure}	
%***************************************************************************

\begin{figure}[!htbp]
\centering
\includegraphics[width=10cm]{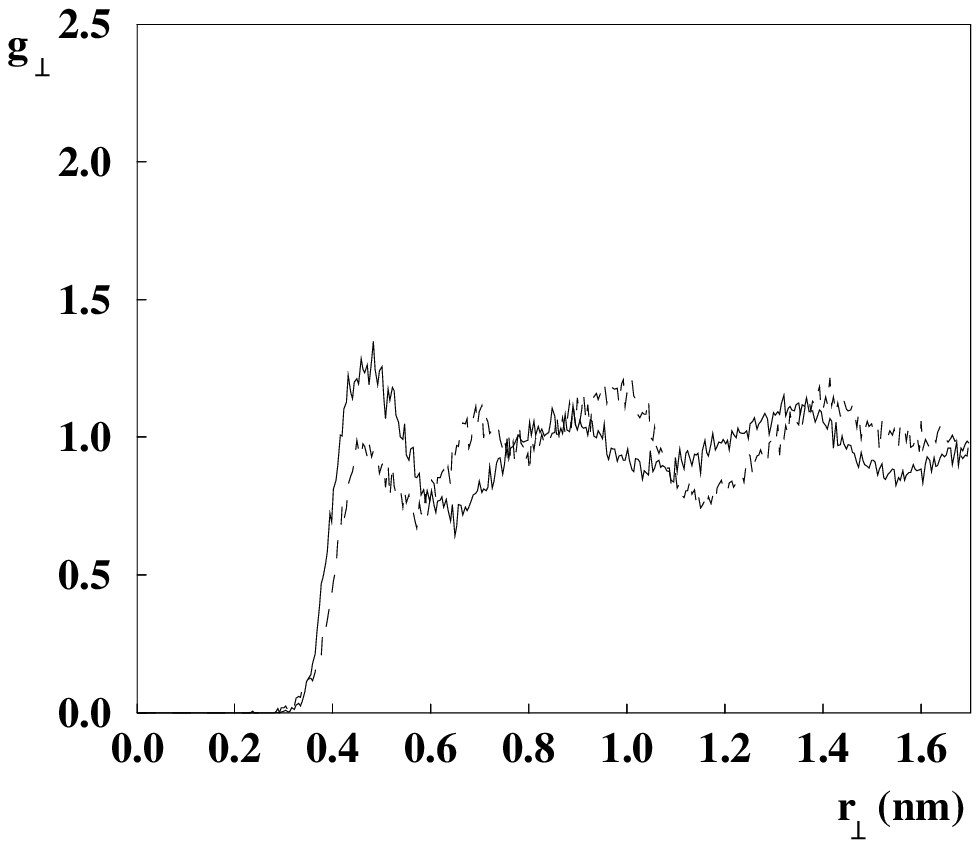}
\caption{The $g_{\perp}$ pair distribution function of C10 atoms
for the system $BM(S)$ at $T$=150~K and $\mu$=1 (solid
line), $\mu$=0.83 (dashed line).}
\protect\label{fig:gperpheads}
\end{figure}

In  Fig.~\ref{fig:gperpdgl} the $g_{\perp}$ distribution of C10 
(panel (a)) and C24 pairs (panel (b)) at $T$=325~K and for the three
densities, $\mu$=1, 0.83 and 0.69, are compared.  C10 is the group that
replaces the charged part of the head of the DMPC molecule in our DMMG model,
while C24 is a group located in the deep hydrophobic region. We see a
significant difference in the translational structure between heads and
tails, which is present at all densities: tails (panel (b)) look
definitely more ordered than heads (panel (a)). At $\mu$=1
both tails and heads are ordered (see Fig.~\ref{fig:gperpheads} and 
the solid line in Fig.~\ref{fig:gperpdgl}~(b)): 
the system appears to be in a crystal
phase. Of course ordering of the heads is progressively lost by increasing
the temperature, as seen by comparing Fig.~\ref{fig:gperpdgl}~(a) 
with Fig.~\ref{fig:gperpheads}. At $\mu$=0.83
tails are still ordered but heads shows a structure which looks
more liquid-like. This is very much indicative of a gel phase. Finally at 
the lowest density, $\mu$=0.69, we see a strong disordering of the
tails, owing to molecular upward/downward rotation, with a
consequent tendency to isotropization. Looking at the shape
of the dotted curve in Fig.~\ref{fig:gperpdgl}~(b) and at the $P_1$ and $P_2$
data at $\mu$=0.69 in Tables~\ref{tab:tab5a} and \ref{tab:tab5b}, 
we are thus led to the conclusion that
at this density the system finds itself in a configuration which is much too
isotropic to be possibly interpreted as the liquid-crystal phase of a
bilayer. It should be noticed, however, that tail disordering does not seem
to affect the value of the inter-molecular energy, because, as we noticed
above, the potential energy loss in each layer is compensated by a stronger
interaction energy between the two layers.
%***************************************************************************
%***************************************************************************
%***************************************************************************
\begin{figure}[!htbp]
\centering
\includegraphics[width=10cm]{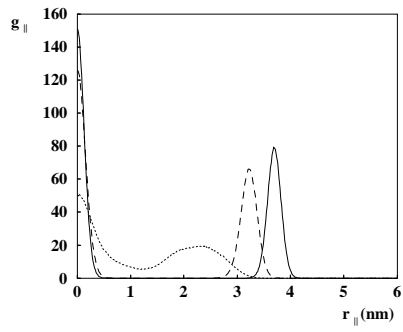}
\caption{The $g_{\parallel}$ pair distribution function of C13 atoms
for the system $BM(S)$ at $T$=325~K and $\mu$=1 (solid line),
$\mu$=0.83 (dashed line), $\mu$=0.69 (dotted line).}
\protect\label{fig:gpar}
\end{figure}	
%***************************************************************************

There is a correlation among the decrease in density, the tilt of the
molecules in the two layers and the decrease in the bilayer thickness. In
Fig.~\ref{fig:gpar} we plot the $g_{\parallel}$ pair distribution functions of
C13 atoms, again for $T$=325~K at the three decreasing densities, $\mu$=1
(solid line), 0.83 (dashed line) and 0.69 (dotted line). The first thing that
should be noted is the presence of very pronounced peaks at $r_{\parallel}$
$\approx$ 0. These peaks come from the contributions of pairs of atoms
belonging to the same layer. Concentrating our attention to the region
$r_{\parallel}>0.5$~nm, we see that the solid and dashed curves show narrow
peaks, sticking out over a region of vanishing $g_{\parallel}$, centered at
$r_{\parallel}\approx 3.8$~nm and $r_{\parallel}\approx 3.3$~nm, respectively.
They are indicative for the existence of two well separated layers. We remark
that the first number is not too far from the value of the experimental
bilayer thickness,  which is $\approx$ 4~nm at the same density and
temperature. The shift in the location of this peak is nicely explained
by the progressive molecule tail tilting, taking place by lowering the
density, which tends to reduce the bilayer thickness. The overall shape of
the dotted curve in Fig.~\ref{fig:gpar} is, instead, completely different:
first of all the rightmost peak is much broader than the corresponding 
peak seen in the
other two curves and secondly there is no region where $g_{\parallel}$ is
zero: in fact, pretty soon, to the left of the rightmost 
peak $g_{\parallel}$ starts to
grow again. This behaviour reflects the fact that at this density the two
layers partially overlap, as a consequence of the tendency to isotropization
described above. Again, we notice that the effect of increasing the
temperature on the bilayer thickness is negligible, if compared to what
happens when we lower the density. For instance, increasing the temperature
from 150~K to 325~K produces a thickness variation (which is maximal at
$\mu$=0.83) of only 0.3~nm, while by decreasing the density from $\mu=1$ to
$\mu=0.83$ and from $\mu=0.83$ to $\mu=0.69$, the peak in Fig.~\ref{fig:gpar}
gets shifted by $\approx$ 0.5~nm and $\approx$ 1~nm, respectively.   
 
\subsubsection{Angular Distributions}
\begin{figure}[!htbp] 
\centering
\includegraphics[width=10cm]{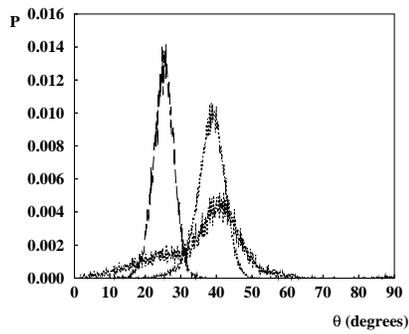}
\caption{The distribution function at $T$=325~K of the angle, $\theta$, 
formed by the segment C34-C46 of tail~\#2 with respect to the bilayer 
normal, for the $BM(S)$ system at $\mu$=1 (long-dashed line), $\mu$=0.83 
(short-dashed line) and $\mu$=0.69 (dotted line).}
\protect\label{fig:dfa}
\end{figure}

%\vskip .2cm  
%\noindent{\it \underline{Angular Distributions}}   
%\vskip .2cm
In Fig.~\ref{fig:dfa} we show the distribution function $P(\theta)$ of the 
polar angle, $\theta$, formed by the bilayer normal with the $z$ axis of the
local frame identified by the atoms C34-C40-C46 of tail~\#2. This direction,
which, we recall, is chosen to be parallel to the vector joining the C34-C46
atoms, is practically aligned with the long molecular axis. We show the
results for the lower layer only, as data for the other layer are
essentially identical. The progressive tail tilting of the molecules with
decreasing density is clearly visible.
%***************************************************************************
%***************************************************************************
We also see that the peak of the distribution moves from 25$^{\circ}$ at
$\mu$=1 (long-dashed line) to 40$^{\circ}$ at $\mu$=0.83 (short-dashed line).
At the lowest density (dotted line) the distribution becomes rather broad, but
still peaked around 40$^{\circ}$. The experimental tilt angle for similar
systems in the gel phase, which has been measured to be about
30$^{\circ}$~\cite{TILT}, is well consistent with these results. We regard
this agreement as a success of our model.

A final structural investigation has been performed on the population of the
different torsional states along the tails. Since the local torsional
potential has three energy wells, corresponding to two {\it gauche}
states ($\phi$=60$^{\circ}$ and 300$^{\circ}$) with the same energy and one
{\it trans} state ($\phi$=180$^{\circ}$) with a different
energy~\footnote{Actually the energy of the {\it trans} state is lower than
the energy in the {\it gauche} states in all torsions involving only
aliphatic carbons.}, we can adequately characterize the situation by a 
parameter, usually called $P_g$, which represents the percentage of the total
{\it gauche} population belonging to each torsion. We find at $T$=325~K the
results reported in Table~\ref{tab:tab6}. In this computation we have
averaged data over the two tails and the two layers. Values smaller than
1~\% are not reported.

\begin{table}[!htbp]
\begin{center}
\caption{Percentage (\%) of population of tail torsions at $T$=325~K
for the $BM(S)$ system at the various densities. Values smaller 
than 1\% are not reported.}
\begin{tabular}{c||ccc} \hline
$Torsion$ & $\mu$=1 & $\mu$=0.83 & $\mu$=0.69 \\
\hline
(C12-C13-O14-C17)+(C13-C15-O16-C32) & 23 & 25 & 24 \\
(C13-O14-C17-C19)+(C15-O16-C32-C34) & 4 & 4 & 7 \\
(O14-C17-C19-C20)+(O16-C32-C34-C35) & 50 & 50 & 44 \\
(C17-C19-C20-C21)+(C32-C34-C35-C36) & 1 & 10 & 7 \\
(C19-C20-C21-C22)+(C34-C35-C36-C37) & - & 2 & 6 \\
(C20-C21-C22-C23)+(C35-C36-C37-C38) & - & 2 & 5 \\
(C21-C22-C23-C24)+(C36-C37-C38-C39) & - & 1 & 5 \\
(C22-C23-C24-C25)+(C37-C38-C39-C40) & - & 1 & 5 \\
(C23-C24-C25-C26)+(C38-C39-C40-C41) & - & 1 & 4 \\
(C24-C25-C26-C27)+(C39-C40-C41-C42) & - & 1 & 5 \\
(C25-C26-C27-C28)+(C40-C41-C42-C43) & - & 1 & 5 \\
(C26-C27-C28-C29)+(C41-C42-C43-C44) &  - & 1 & 6 \\
(C27-C28-C29-C30)+(C42-C43-C44-C45) &  1 & 4 & 8 \\
(C28-C29-C30-C31)+(C43-C44-C45-C46) &  3 & 6 & 9 \\
\label{tab:tab6}
\end{tabular}
\end{center}   
\end{table} 
%******************************************************************** 
At the two highest densities only the last two torsions (C27-C28-C29-C30,
C28-C29-C30-C31) in tail~\#1 and the corresponding ones in tail~\#2 show a
population larger than 1\%. The high values, $P_g$=25\% and $P_g$=50\%, that
we find for the average percentage population in the torsions
(C12-C13-O14-C17) + (C13-C15-O16-C32) and (O14-C17-C19-C20) +
(O16-C32-C34-C35), respectively, are due to the distortion of the geometry
in the ``kinked" region close to the ester bond. These values are
essentially the same for all densities and temperatures and are due to the
phenomenon of ``kink inversion" between the two tails, described before. It
is worth noticing that the $\mu$=0.83 results are consistent with those
obtained for the DPPC in the gel phase~\cite{K,KLEIN_DPPC}, after
correcting for the fact that the DPPC hydrophobic tails are 2 methylene
groups longer than the DMPC tails.

The smooth increase of the {\it gauche} population along the tails, seen 
in the simulations performed at the lowest density, means that the
rotational freedom of molecule tails does not help in making the tail
flexibility in the terminal region larger than that in the core. This is at
variance with the results of the first paper of ref.~\cite{KLEIN_DPPC}, 
where a larger mobility of
tail end segments, as compared to what is seen in the core region, was found
in simulations of the liquid-crystal phase of DPPC. Values around 20~\% for
$P_g$ for the last six torsions of the tails are, in fact, reported. One
might argue that the little tail flexibility we find in our model might be
attributed to the much too strong 1-4 LJ potential we have
introduced in each torsion. These 1-4 forces mainly have the effect of adding
a steric repulsion in the {\it gauche} conformations. We recall that
within the framework of
the OPLS force-field~\cite{OPLS}, which we employed in the present paper, we
have reduced, according to the general OPLS philosophy, the strength of the
1-4 LJ potential by a factor 8. This may not have been enough in
view of the too high tail rigidity we have found. However, neglecting 1-4
forces altogether gives results only partially in the desired direction. In
fact, on the one hand it actually produces an increase of the {\it gauche}
population by a large factor (in trial simulations performed on liquid
butane this factor is found to be about 3), but on the other hand it leads
to a more pronounced upward/downward molecular rotational freedom, which may
make even more problematic for the system to unveil its liquid-crystal
phase.    

\subsubsection{Segmental Order Parameter}
\begin{figure}[!htbp]   
\centering 
\includegraphics[width=10cm]{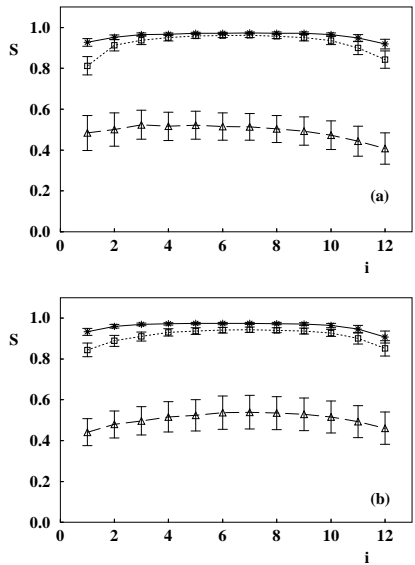}
\caption{The $S$ order parameter for the system $BM(S)$
at $T$=325~K as a function of the position
along tail~\#1 (a) and tail~\#2 (b): stars are for
$\mu$=1, squares for $\mu$=0.83 and triangles for $\mu$=0.69. The first
segment starts from the carbonyl carbon atom C17 in tail~\#1 and C32 in
tail~\#2. The lines joining the points are only to guide the eye.}  
\protect\label{fig:po1}
\end{figure}

%\vskip.2cm   
%\noindent{\it\underline{Segmental Order Parameter}}  
%\vskip .2cm 
In Fig.~\ref{fig:po1} we show the behaviour of the order parameter, 
$S_i$ (eq.~(\ref{OP}) of sect.~3.1), as function of the position of the
(C$_i$-C$_{i+2}$) segment along the tail~\footnote{We observe that the
notation (C$_i$-C$_{i+2}$) does not apply to the initial segments of the
two tails, that are called C17-C20 and C32-C35 respectively, because in the
enumeration of carbons the labels 18 and 33 are attributed to O atoms.}.
Panels (a) and (b) show the results for tail~\#1 and~\#2, respectively,
averaged over the two layers. We focus on the behaviour of $S_i$ at
$T$=325~K and different densities. We see that, correctly, the order
parameter decreases with the density (the expectation for a completely
isotropic fluid would be $S_i$=0). Notice that, not surprisingly, data at
the lowest density are affected by errors that are definitely larger than
those on the data at $\mu$=1 and $\mu$=0.83.
%***************************************************************************
%***************************************************************************
%***************************************************************************

\subsection{The system $BM(L)$}

As summarized in Table~\ref{tab:tab2}, we have performed extensive MD
simulations also on the much larger system, $BM(L)$, consisting of 2 layers
of 256 DMMG molecules each. To compare with previous results we have analyzed
the trajectories corresponding to the same combined values of surface
density and temperature considered above. An immediate observation that
emerges from the analysis of $BM(L)$ MD data is that statistical errors are
about the same as those one finds in the case of the system $BM(S)$, despite
the fact that substantially shorter trajectories were produced (compare 
Tables~\ref{tab:tab1} and~\ref{tab:tab2}). As we already noticed, this is a
direct consequence of the increase in the size of the system and of the fact
that the larger $BM(L)$ system was built by assembling already equilibrated
$BM(S)$ blocks with the result that no special needs for further long or
sophisticated equilibration procedures have emerged.

The general conclusions of the analysis of $BM(L)$ data is that the scenario
we have come up with in the previous sections is well confirmed by the
simulations of this larger system, and we have good consistency between the 
two sets of data within errors.

%***************************************************************************
\begin{figure}[!htbp]
\centering
\includegraphics[width=10cm]{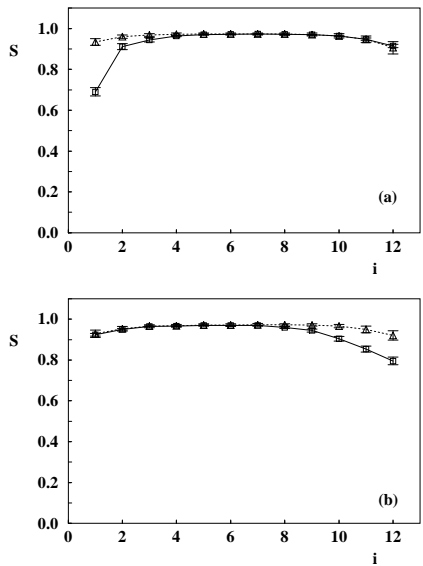}
\caption{ The $S$ order parameters at $T$=325~K and $\mu$=1 
as function of the position along tail~\#1 (a) 
and tail~\#2 (b) for the $BM(L)$ system (squares)  and for
the $BM(S)$ system (triangles). The lines joining the points are only 
to guide the eye.}
\protect\label{fig:po2}
\end{figure}
\begin{figure}[!htbp] 
\centering 
\includegraphics[width=10cm]{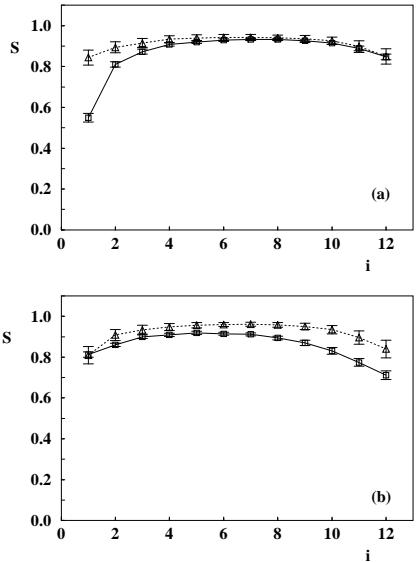}
\caption{ Same as in Fig.~\ref{fig:po2} at $\mu$=0.83.}
\protect\label{fig:po3}
\end{figure}

	
%***************************************************************************
As an example of the quality of the results that we have obtained, we compare
in Figs.~\ref{fig:po2} and~\ref{fig:po3} the behaviour of the order parameter
$S$, for the two systems, $BM(S)$ and $BM(L)$ at $T$=325~K and $\mu$=1 and
$\mu$=0.83, separately for the two tails, averaged over the two layers. The
$BM(L)$ points are systematically lower than the $BM(S)$ data, as expected
from the lowering of the collective orientational order with the increasing
size of the system. This effect is more relevant for segments close to the
head and to the end of the tails. The slightly larger errors that affect the
data associated to the first and the last tail segment are due to jumps among
different torsional states that are not sufficiently well sampled 
in the time evolution
at our disposal. In order to reduce errors associated to this slow torsional
dynamics, it would be important to increase the length of $BM(L)$ simulations.
Data at $\mu$=0.69 are affected by substantially larger errors and are not
shown.
%***************************************************************************
%***************************************************************************

The above indications about the behaviour of the degree of collective
orientational order with the size of the system are confirmed by a careful
study of the order parameters, $P_1$ and $P_2$. For instance, at $\mu$=0.83 
and $T$=325~K one gets for the $BM(L)$ system $P_2$=0.964 $\pm$ 0.003, a 
value somewhat lower than the one observed for the smaller $BM(S)$ system,
for which an average value $P_2$=0.984 $\pm$ 0.002 is found. A lowering of
the orientational order with the size of the system is actually expected
from the observation that in a bigger system fluctuations with longer wave
lengths are possible. 

Further agreement among $BM(L)$ and $BM(S)$ simulations emerges from the 
study of translational order properties. The $g_{\perp}(r)$ distribution
functions computed from the $BM(L)$ simulations show essentially the same
structure and locations of the peaks that is seen in the $BM(S)$
distributions. The only difference is a somewhat less pronounced peak
structure in distributions referring to atoms belonging to the hydrophobic
core. Given the strong similarity of $BM(L)$ and $BM(S)$ results, we do not
report the former here.

We would like to conclude this section with some general considerations
about the quest for MD simulations of larger and larger systems. Given the
satisfactory level of consistency we find by comparing the MD data of the two
samples, one might ask whether simulations of systems as large as $BM(L)$
are really necessary to get the physics of the system right. Although the
answer to this question strongly depends on the kind and the degree of detail
of the information one wishes to obtain from the model, it is important to
recall that the study of finite volume effects is a necessary step in any
numerical simulation, because features and properties seen at small volumes
may completely change when the volume is increased. Consequently it is
absolutely mandatory to get an idea of the magnitude of finite volume
corrections. From our experience it was only thanks to the data we had for
the larger system that we could reliably draw useful conclusions on the
structural properties of our model-membrane.   

Apart from these rather general considerations, there are in our opinion
many specific problems where the fact of having a sufficiently large system
is crucial in order to be in the position of extracting reliable results from
numerical simulations. Just to make two examples we may mention:

1) the problem of trying to accurately locate the critical lines of the phase
diagram of the system. The obvious reason for the necessity of comparing
systems of different sizes is that a structural change of the system can be
interpreted as a true phase transition only if one is able to show that the
transition becomes sharper and sharper as the volume increases. 

2) Another very interesting problem is the study of the dynamic and the
formation of ion channels. In this case dealing with a sufficiently large
system is required by the need of having the cross section of the channel 
sufficiently smaller than the surface of the bilayer itself in order to meet
as accurately as possible the actual physical situation. We wish to mention
in this context the pioneering investigation recently carried out in the
beautiful work of ref.~\cite{DIECK}, 
where it has been shown that indeed channels
have the possibility of forming and dynamically maintaining themselves
stable.

\section{Conclusions}

We have performed a systematic MD study of the structural properties of a
simple atomistic model of bilayer by analyzing the behaviour of a large
number of physical quantities characterizing the system, as functions of
density and temperature, and by comparing results obtained from simulations
carried out on two systems of largely different size.

We have found that our modelization reproduces in a remarkable way many known
features of the gel phase of DMPC bilayers~\cite{SSSC}, as well as the
structural changes that accompany the transition from the crystal to the gel
phase. These findings have emerged quite clearly by analyzing the very long
trajectories we have collected for the system, $BM(S)$, and are nicely
confirmed by the statistically equally accurate data we have for the much
larger system, $BM(L)$.

Clearly a lot of work is still needed to further validate and improve our
approach, especially in the direction of investigating the role of water,
whose effects were only indirectly accounted for through the modulation of the
surface density of the system. The inability of the model to cope with the
expected liquid-crystal phase of the system at low density is its major
limitation. To prevent the extrusion of the molecule tails from the planes of
the bilayer at low density and high temperature, either some sort of
geometric or dynamic constraint must be introduced, perhaps by modifying the
intra-molecular tail flexibility, or more directly one should explicitly
introduce water. 

Despite the limitations we have mentioned, from the quality of the results
presented in this investigation, we conclude that the membrane model we have
developed is capable of capturing most of the essential structural
properties of the real system in the crystal and gel phase. Thanks to its
simplicity, we are confident that it will be possible to appropriately
improve the model in the directions mentioned above, without affecting the 
key features that are at the basis of the stability of our bilayer, {\it
i.e.}~tail packing within the layers and adequately strong interactions
between the two layers.

{\bf Acknowledgments} - We wish to thank G.~Ciccotti and S.~Melchionna for
discussions. We also thank the ENEA Computing Center (Casaccia - Italy) 
and CINECA (Bologna - Italy), where most of the computations presented here
have been performed and the APE groups of the Universities of Rome {\it
{La Sapienza}} and {\it {Tor Vergata}} for their assistance. Partial
support from the italian Institutions INFN, INFM, CNR and MURST is
gratefully acknowledged.

\end{document}